\documentclass[article]{agujournal2019}
\usepackage{url} 
\usepackage{lineno}
\usepackage[finalnew]{trackchanges} 
\usepackage{soul}

\usepackage{dcolumn}
\usepackage{graphicx}
\usepackage{color}
\usepackage{amssymb}
\usepackage{bold-extra}
\usepackage{caption}

\def\figref#1{Fig.~\ref{fig:#1}}
\def\figlab#1{\label{fig:#1}}  
\def\eqref#1{Eq.~(\ref{eq:#1})}

\newcommand*{\secref}[1]{Section~\ref{sec:#1}}
\newcommand*{\seclab}[1]{\label{sec:#1}}

\RequirePackage[normalem]{ulem}

\draftfalse
\journalname{JGR: Atmospheres}

\begin{document}

\title{The initial stage of cloud lightning imaged in high-resolution}

\authors{O.~Scholten\affil{1,2,3}, B.~M.~Hare\affil{1},
J.~Dwyer\affil{4}, C.~Sterpka\affil{4},
I.~Kolma\v{s}ov\'{a}\affil{5,6}, O.~Santol\'{i}k\affil{5,6}, R.~L\'{a}n\affil{5}, L.~Uhl\'{i}\v{r}\affil{5},
S.~Buitink\affil{7,8}, A.~Corstanje\affil{7,8}, H.~Falcke\affil{7,9,10,11}, T.~Huege\affil{8,12}, J.~R.~H\"orandel\affil{7,8,9}, G.~K.~Krampah\affil{8}, P.~Mitra\affil{8}, K.~Mulrey\affil{8}, A.~Nelles\affil{13,14}, H.~Pandya\affil{8}, A.~Pel\affil{2}, J.~P.~Rachen\affil{8}, T.~N.~G.~Trinh\affil{15}, S.~ter Veen\affil{10}, S.~Thoudam\affil{16}, T.~Winchen\affil{8}}

\affiliation{1}{University Groningen, Kapteyn Astronomical Institute, Landleven 12, 9747 AD Groningen, The Netherlands}
\affiliation{2}{University Groningen, KVI-Center for Advanced Radiation Technology, P.O. Box 72, 9700 AB Groningen, The Netherlands}
\affiliation{3}{Interuniversity Institute for High-Energy, Vrije Universiteit Brussel, Pleinlaan 2, 1050 Brussels, Belgium}
\affiliation{4}{Department of Physics and Space Science Center (EOS), University of New Hampshire, Durham NH 03824 USA}
\affiliation{5}{Department of Space Physics, Institute of Atmospheric Physics of the Czech Academy of  Sciences, Prague, Czechia}
\affiliation{6}{Faculty of Mathematics and Physics, Charles University, Prague, Czechia}
\affiliation{7}{Department of Astrophysics/IMAPP, Radboud University Nijmegen, P.O. Box 9010,  6500 GL Nijmegen, The Netherlands}
\affiliation{8}{Astrophysical Institute, Vrije Universiteit Brussel, Pleinlaan 2, 1050 Brussels, Belgium}
\affiliation{9}{NIKHEF, Science Park Amsterdam, 1098 XG Amsterdam, The Netherlands}
\affiliation{10}{Netherlands Institute of Radio Astronomy (ASTRON),  Postbus 2, 7990 AA Dwingeloo, The Netherlands}
\affiliation{11}{Max-Planck-Institut f\"{u}r Radioastronomie, P.O. Box 20 24,  Bonn, Germany}
\affiliation{12}{Institut f\"{u}r Kernphysik, Karlsruhe Institute of Technology(KIT), P.O. Box 3640, 76021, Karlsruhe, Germany}
\affiliation{13}{Erlangen Center for Astroparticle Physics, Friedrich-Alexander-Univerist\"{a}t Erlangen-N\"{u}rnberg, Germany}
\affiliation{14}{DESY, Platanenallee 6, 15738 Zeuthen, Germany }
\affiliation{15}{Department of Physics, School of Education, Can Tho University Campus II, 3/2 Street, Ninh Kieu District, Can Tho City, Vietnam}
\affiliation{16}{Department of Physics, Khalifa University, PO Box 127788, Abu Dhabi, United Arab Emirates.}

\correspondingauthor{Olaf Scholten}{O.Scholten@rug.nl}

\begin{keypoints}
\item Our new LOFAR imaging procedure can locate over 200 sources per millisecond of flash with  meter-scale accuracy.
\item The Primary Initial Leader breaks up into many (more than 10) negative leaders of which only one or two continue after 30 ms.
\item Some negative leaders propagate from the positive charge layer back to get close to the initiation point.
\end{keypoints}

\begin{abstract}
With LOFAR 
we have been able to image the development of lightning flashes with meter-scale accuracy and unprecedented detail.
We discuss the primary steps behind our most recent lightning mapping method. To demonstrate the capabilities of our technique we show and interpret images of the first few milliseconds of two intra-cloud flashes. In all our flashes the negative leaders propagate in the charge layer below the main negative charge. Among several interesting features we show that in about 2~ms after initiation the Primary Initial Leader triggers the formation of a multitude (more than ten) negative leaders in a rather confined area of the atmosphere.   From these only one or two continue to propagate after about 30~ms to extend over kilometers horizontally while another may propagate back to the initiation point.
We also show that normal negative leaders can transition into an initial-leader like state, potentially in the presence of strong electric fields. In addition, we show some initial breakdown pulses that occurred during the primary initial leader, and even during two "secondary" initial leaders that developed out of stepped leaders.
\end{abstract}



\section{Introduction}

One of the key open questions in lightning science concerns the understanding of the processes that are fundamental to the initiation and early development of a lightning flash. In particular, it is not known what processes lead to the creation of the primary initial leader (PIL) channel and how that channel propagates. In recent years, the use of lightning mapping arrays \cite{Rison:1999, Edens:2012} and VHF radio interferometers \cite{Rhodes:1994, Yoshida:2010, Stock:2014} augmented with fast antennas and optical measurements \cite{Hill:2011, Montanya:2015, Qi:2016, Tran:2016} have led to the general picture that lightning initiation begins with an ionization event \cite{Stolzenburg:2020}, which could be in the form of a powerful narrow bipolar event \cite{Rison:2016} or much weaker VHF source as seen in \cite<e.g.>{Marshall:2019, Lyu:2019}. This is followed by an initial leader, as imaged very nicely in \cite{Lyu:2016} in VHF.  The initial leader propagation usually involves a series of large preliminary breakdown pulses (see for example \cite{Kolmasova:2014, Kolmasova:2018}), after which normal negative stepped leader propagation is observed. The transition from the initial leader to a negative stepped leader has been observed in \cite{Stolzenburg:2020} with high-speed video and in electric field change data. Why negative leaders initially propagate in a different mode than the normal leader stepping seen later in the flash is not understood. Furthermore, the positive leader is often not observed during the initiation process and only appears in radio data much later on after the negative leader is well developed.

With the present work we add very accurate images, obtained using LOFAR, showing the dynamics of Dutch thunderstorms. LOFAR \cite{Haarlem:2013} is a software-phased array consisting of several thousand simple antennas that is primarily built for radio astronomy, see \secref{LOFAR}. Thunderstorms we have observed in the general area of the Dutch LOFAR stations (Dutch thunderstorms) differ from the thunderstorms seen in the US by the fact that all the flashes we have observed initiate at the bottom of the main negative charge layer, and then propagate down into the lower positive charge layer. For many flashes, including the two discussed more extensively in this work, we observe an extensive network of negative leaders which very rarely result in a ground stroke. It thus appears that the negative leaders become ``trapped" in the potential well of the lower positive charge layer.
To improve the insight in the dynamics of the lightning discharge immediately after initiation we have improved our imaging technique over our earlier procedure \cite{Hare:2019} by greatly increasing the number of located VHF sources. In \secref{Imaging} we elucidate on the main improvements of our present imaging procedure.

Some detailed images of the initial stage of the lightning discharge are shown in \secref{Initiation}, partially to demonstrate the capabilities of our present imager. We show images for two flashes, one from 2018 and one from 2019. The 2018 flash shows the typical initial development we have seen in all our imaged flashes, about ten in number. The flash starts with a very small pulse in VHF, barely recognizable even with our sensitive LOFAR antennas. This develops into a Primary Initial Leader, as also imaged in \cite{Lyu:2016}. During its development we detect rapidly increasing VHF (30 -- 80 MHz) activity reaching a maximum about 2~ms after initiation. For the 2019 flash we also have data recorded from a broadband  magnetic-loop antenna during this time, showing significant low-frequency emissions at particular stages of the PIL development. After descending down from the negative to the positive charge layer the PIL initiates a plethora of negative leaders almost simultaneously in an area of about 1~km$^2$. Of the original multitude of negative leaders, only one or two continue to propagate after about 30~ms to form the main part of the flash which may cover distances of 10 km or more. The 2019 flash is interesting since we see there a PIL and even two secondary Initial Leaders (SIL). The Initial Leaders are clearly distinguishable from a negative leader through their propagation speed, a relatively low density of imaged sources and powerful VHF emission.

Some suggestions for a possible interpretation of our observations are presented in \secref{discuss}.

\section{Methods}

Our mapping procedure of pulses detected by LOFAR basically follows the structure outlined in \cite{Hare:2019}. Arrival-time differences for pulses coming from the same source in different antennas are extracted from the data using the maxima in the cross correlations.
The present procedure incorporates important improvements that are mainly due to an improved procedure, inspired by that of the Kalman filter, to follow the pulse from the same source across many different antennas that may be many tens of kilometers apart. For completeness we outline here the procedure we followed \cite{Scholten_LofarImaging:2020}.

\subsection{LOFAR}\seclab{LOFAR}

\begin{figure}[h]
\centerline{\includegraphics[width=0.49\textwidth]{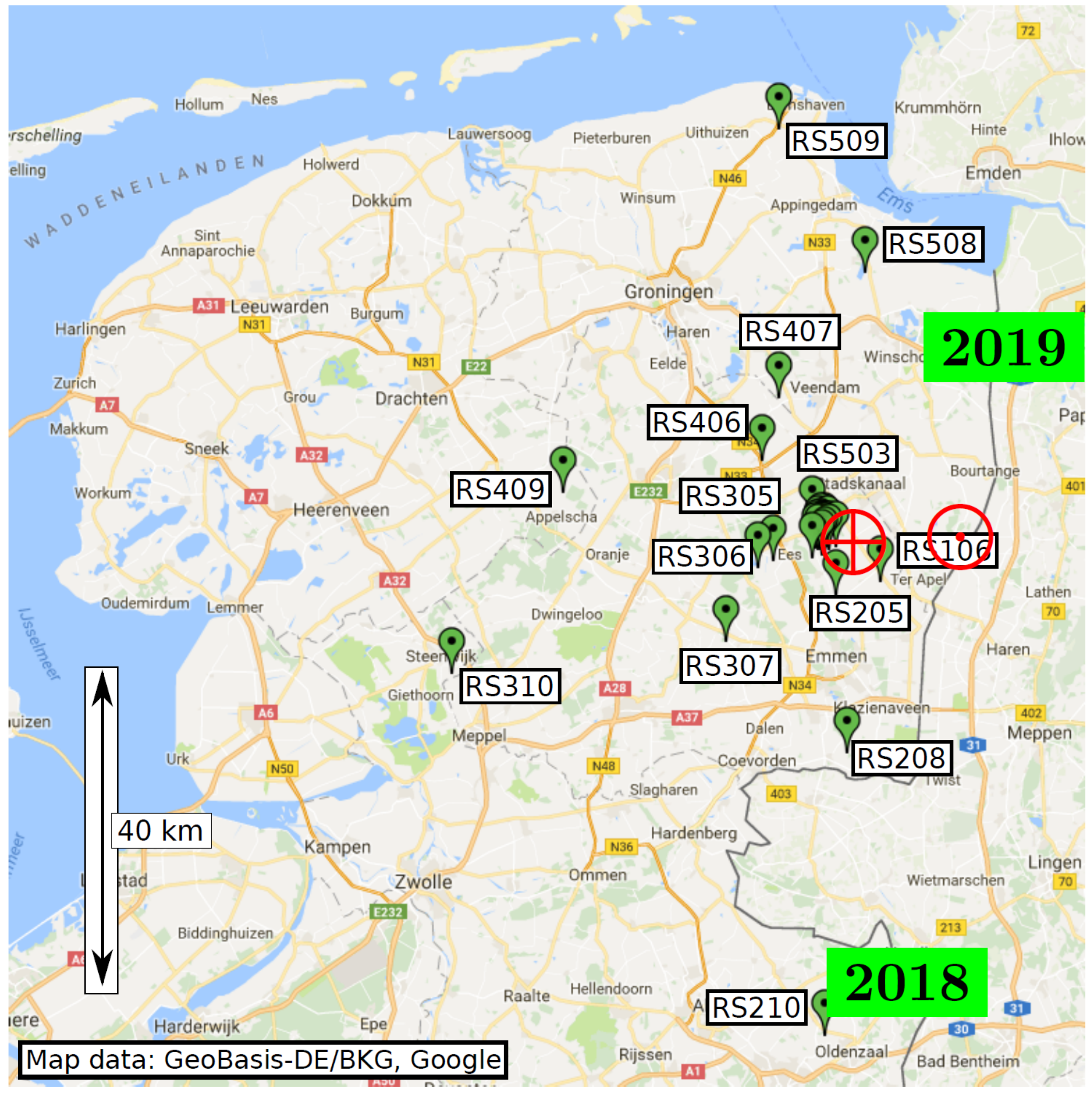} }
	\caption{Layout of the Dutch LOFAR stations, adapted from \cite{Hare:2018}. The core of LOFAR is indicated by the red $\bigoplus$\ sign, the position of the broadband magnetic loop antenna SLAVIA 
is indicated by the red $\bigodot$, while the green boxes show the general location of the 2018 and 2019 flashes that are discussed in this work.}
	\figlab{LOFAR-NL}
\end{figure}

LOFAR \cite{Haarlem:2013} is a radio telescope consisting of several thousands antennas. These antennas are spread over a large area with a dense core (a circular area with a diameter of 300~m), the Superterp, near Exloo, the Netherlands, and with remote stations spread over Europe, reaching baselines in excess of 1000~km. The signals from these antennas can be added coherently to make this effectively operate as a gigantic radio dish, primarily used for radio astronomy.
For our lightning observations we confine ourselves to the LOFAR stations in the Netherlands, reaching baselines of the order of 100~km, see \figref{LOFAR-NL}. We use the Low Band Antennas (LBAs) operating in the frequency range from 30 -- 80~MHz. The LOFAR antennas are arranged in stations. Each station has 96 dual polarized antennas with an inverted V-shape. The signals are sampled at 200~MHz (5~ns sampling time). For our observations we use about 12 antennas (6 for each polarization) per station. For the lightning observations the circular memory (called Transient Buffer Board, TBB) is used that can store 5~s of data per antenna. Upon an external trigger, taken from \cite{Blitzortung}, the data on the TBBs are frozen and read out for later processing. In this read-out process we experience some data loss (due to missed hand-shaking during the download from the antenna field) which, thanks to our large number of antennas, does not affect the image quality. Per 5~s recording we store close to 1~TB of data for later off-line processing. The antennas have been calibrated on the galactic background radiation \cite{Mulrey:2019}.

\subsection{Lightning Imaging}\seclab{Imaging}

The basis of our imaging procedure is described in \cite{Hare:2019}. We choose a reference antenna which usually is located in the core at the Superterp. For each pulse for which we want to search for the source location we select a relatively small section of the time trace in the reference antenna. The arrival times of pulses from this source in other antennas are calculated from the cross correlation of the selected trace with the traces in the other antennas. The source position is determined from a chi-square fit of these arrival times.

Imaging a flash starts with RFI mitigation, see \secref{RFI}. The time calibration of all participating antennas, discussed in \secref{Calibr} is the step that requires most attention since we want to reach an accuracy of 1~ns for all antennas. Finding the source positions is the third step which is through a fully automatized pipeline, see \secref{SourceFind}. For the final image we select those sources that obey certain quality conditions, see \secref{SourceQ}, where one has to balance keeping a sufficient number of sources with limiting the scatter. Full details of the new procedure are given in \cite{Scholten_LofarImaging:2020}, here we will outline the main aspects.

\subsubsection{RFI mitigation}\seclab{RFI}

Since the LOFAR core is situated in a rather densely populated part of the world there are many radio and TV transmitters that interfere with our observations. Because of these our frequency range is limited to 30 -- 80~MHz. At lower and higher frequencies there is too much RFI to excise it. The few narrow-frequency lines in our detection window we mitigate by software notch filters.

\subsubsection{Timing calibration}\seclab{Calibr}

Since we want to achieve meter-scale resolution the relative timing of the antennas has to be calibrated at the nanosecond level. To achieve this over distances of 100~km we use a few selected pulses emitted during the flash in a bootstrap procedure. Regularly spread over the duration of the flash a small number (order of five) blocks of data (one block is 32k of 5 nanosecond time samples) are taken for the reference antenna, which is taken in the dense core of LOFAR on the Superterp.

Here, and later in the imaging procedure, we minimize, using a Levenberg-Marquardt algorithm, the root mean square time difference ($RMS$) between the calculated arrival times and measured arrival times for all antennas to find the position of a source. The calculation uses the travel time of a signal from the source to the antenna. The  measured arrival times are obtained from the peak position in the absolute value of the cross correlation between a small part around the pulse in the reference antenna and the spectrum in the antenna.

In each block, up to the four strongest pulses are identified as candidate calibration pulses. The pulses from the same candidate sources are selected in the nearby stations. A candidate is eliminated from the calibration procedure if there is an ambiguity in selecting the correct pulse in the adjacent antennas.
The known LOFAR timing calibrations are sufficient at this stage which for the core has an accuracy better than 5~ns. The $RMS$ is minimized to find the locations of the candidate calibration sources.
The Dutch antennas are separated in rings with increasing diameter. In iterations that follow, a larger ring of antennas is included. For each iteration the source locations found in the previous iteration is used to make an educated guess of the pulse timings for all antennas from these calibration sources. In a chi-square fit, the optimal station timing calibrations (the same for all pulses) are determined while at the same time updating the source locations. A station is excluded from the procedure for a particular calibration source when the pulses attributed to this source  show a large difference with the actual arrival times for all antennas in this station.

The fitting procedure is repeated increasing the radius of the ring around the reference antenna until all stations are included. Frequent visual inspection of the cross correlation spectra is important to guarantee that the used pulses are correctly assigned to the correct calibration  source. When there is doubt, the candidate calibration source is eliminated from the procedure.

In the final stage, the antennas in each station are calibrated (allowing for differences between antennas in the same station) by fitting simultaneously the antenna timings as well as the source locations of the remaining high-quality calibration sources, taking the previously obtained results as an initial guess.

\subsubsection{Source finding}\seclab{SourceFind}

The general source-finding stage is usually run as a standalone process.  This is in contrast to the calibration stage, which requires human inspection. The time trace for the whole flash is divided into overlapping blocks (32k of 5 nanosecond time samples) for each antenna and further processing is done on each block. The overlap is chosen such that the pulses from sources anywhere in the general area of the flash can be recovered in all antennas.

The block of the reference antenna is searched for candidate pulses to be imaged. The overlap regions are excluded so that the same source is not imaged twice.
Candidate pulses are the strongest ones that differ in peak position by more than about 100~ns (the exact time difference depends on the width of the pulse) in the reference antenna and are seen in the two antenna polarizations (dual polarization). Within each block the candidate pulses are ordered in decreasing peak amplitude.

For each candidate pulse in the reference antenna, a section of the time trace around the pulse is taken for the calculation of the cross correlation with other antennas.
Imaging (source finding) starts by performing a grid search over source locations for minimizing the $RMS$ for antennas on the Superterp. The thus obtained source location is taken as the initial guess for a chi-square search to find the optimal source location by minimizing the $RMS$ for all antennas within a certain distance from the core.

This fitting is repeated for an increasing number of antennas by increasing the circle of included antennas. Peaks in the cross correlations are searched for within a timing window that is calculated from the covariance matrix that was obtained from the previous chi-square fit. If the peak in the cross correlation deviates by more than two standard deviations the antenna is flagged as excluded from the fit. An antenna is also excluded when the width of the cross correlation, defined as the integral divided by the peak value, differs by more than 60\% from that of the self correlation in the reference antenna.
The reason for excluding antennas is that it may happen that two pulses are close, or even interfere, for a particular antenna. It may also happen that the pulse is 'hidden' in the noise. Not capturing this may derail the search for the source location.

The procedure of finding the source by gradually including more antennas while limiting the search window is inspired by that of the Kalman filter however is more accurate than even the extended Kalman filter, see \cite{Pel:2019} for an implementation of the Kalman filter for lightning imaging.

After finding the source location the corresponding locations in the trace of each antenna is set to zero and the following candidate pulse is taken.

The inverted V-shaped LOFAR antennas have two possible orientations, SW-NE and SE-NW, and are thus sensitive to different polarizations of the incoming radiation. We notice that the pulse-shape may differ for the two polarizations, see \secref{2018-pulse} for an example. For this reason we have organized our imaging algorithm such that only one of the antenna polarizations or both can be used in imaging where dual (both) is the default.

\subsubsection{Source quality}\seclab{SourceQ}

We observe that the imaging accuracy of a source is poorly reflected by the covariance matrix that is obtained from minimizing the $RMS$. The reason is that selecting a wrong pulse in a series of antennas may still yield a reasonable fit but will result is a source that is mislocated. We have observed that the obtained value for the $RMS$ combined with the number of excluded antennas, $N_{ex}$, appear to be good additional indicators of the image quality supplementing the diagonal element of the covariance matrix corresponding to the error on the height,  $\sigma(h)^2$, which is usually the largest.

Antennas are excluded from the fitting procedure when there is no clear peak in the part of the spectrum that was searched. This could be due to the fact that the pulse is simply too weak to be seen but it could also be that the peak lies outside the search window. The latter is obviously problematic and should have contributed to the $RMS$.
The setting of the imaging quality indicators ($\sigma(h)$, $RMS$, and $N_{ex}$) is dependent on the location of the flash with respect to the core. Additionally the criteria tend to be subjective, balancing a large number of sources with a minimum of mislocated sources. In many cases the mislocated sources appear to be displaced by 50 meters or more along the radial direction with the core of LOFAR at the center.

\subsection{Broadband measurements with a magnetic loop antenna}\seclab{ML}

In September 2018, the broadband magnetic loop antenna SLAVIA (Shielded Loop Antenna with a Versatile Integrated Amplifier) has been installed  by the Department of Space Physis, Institute of Atmospheric Physics of the Czech Academy of Sciences at a site about 10~km east from the LOFAR core near the village Ter Wisch, marked with the red $\bigodot$ in \figref{LOFAR-NL}. The antenna  has a surface area of 0.23~m$^2$ and measures the time derivative of the magnetic field. The obtained  waveforms are then numerically integrated. The sampling frequency is 200~MHz, the frequency band is limited by a first order high-pass filter at 4.8~kHz and by a 13$^{th}$ order low-pass filter at 90~MHz. The sensitivity of the recording system is 6~nT/s/$\sqrt{\rm Hz}$ corresponding to 1~fT/$\sqrt{\rm Hz}$ at 1~MHz. At the Ter Wisch site, the signal is unfortunately affected by strong man-made interferences, some of which cut through our high pass filter, and the waveform had to be cleaned by \remove{ten narrow-band filters at interference frequencies around 2, 2.5, 2.9, 4.2, 5, 5.5, 6.1, 7, 8, and 10~kHz.} \add{19 narrow band-rejection filters with bandwidths 18-30 Hz at interference frequencies between 2 and 10 kHz, and at 18 kHz.}

\section{The initial stages of Dutch lightning flashes}\seclab{Initiation}

In this work we concentrate on imaging the initial development of two lightning flashes, where we almost randomly selected one from 2018 and another one from 2019.
The 2018 flash shows features we see in all our imaged flashes (about 10 in total). A PIL is initiated at the lower side of a negative charge layer. This Initial Leader propagates with a velocity of about $10^6$~m/s downward to the positive charge layer where it simultaneously initiates many negative leaders of which one or two continue to propagate over large distances. The 2019 flash shows a more complicated pattern which can be understood as a PIL initiating in the usual way several negative leaders of which two convert into an initial leader again after a few milliseconds. This second generation of initial leaders we have named Secondary Initial Leaders. In \secref{2019} we will also present data for the 2019 flash from a broadband magnetic loop antenna \cite{Kolmasova:2018}.

\subsection{The 2018 flash}\seclab{2018}

\begin{figure}[h]
\centering	\includegraphics[bb=0.5cm 2cm 22.5cm 27cm,clip, width=0.89\textwidth]{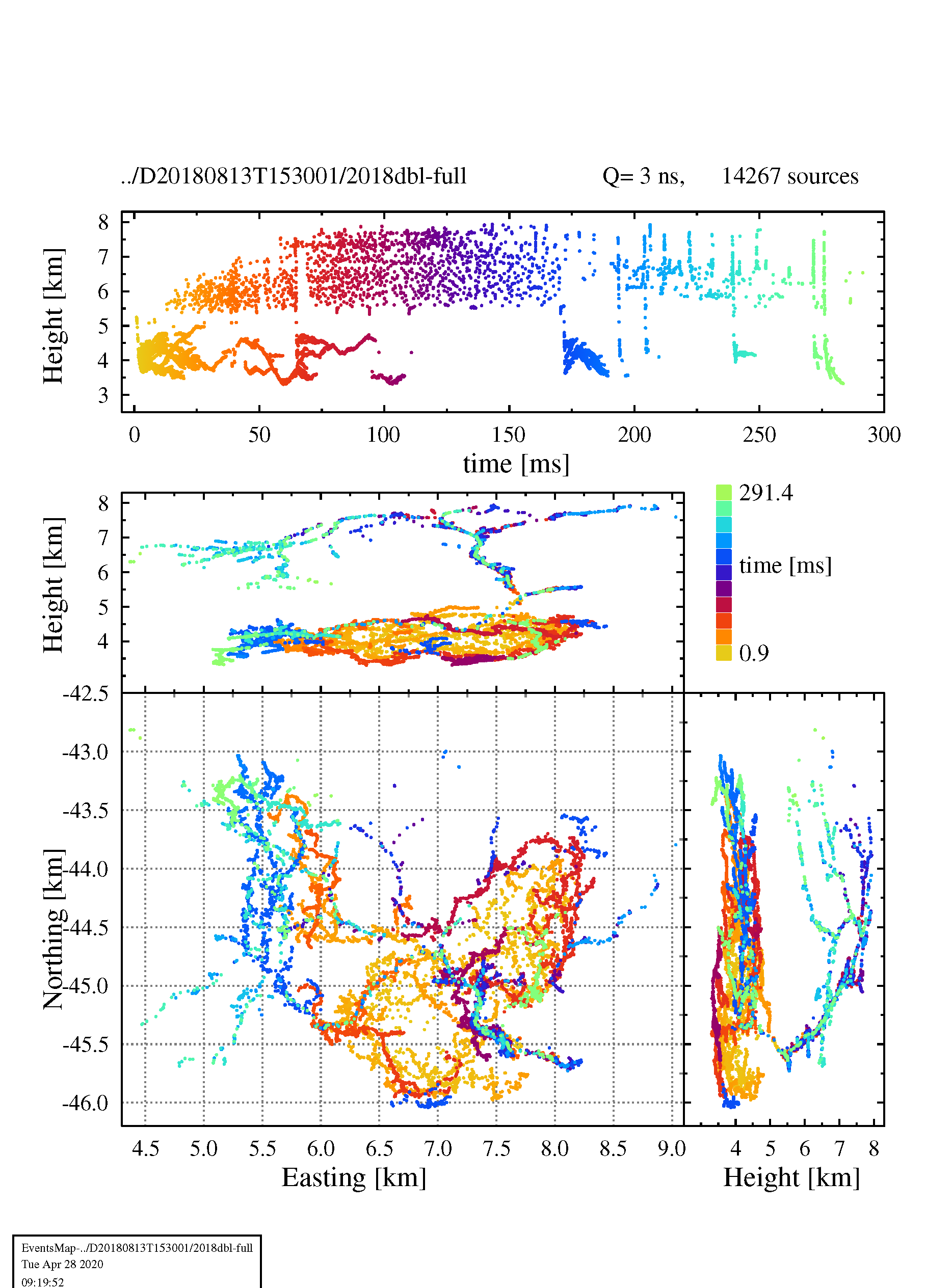}
	\caption{Image of the 2018 flash showing sources with $\sigma(h)<3.5$~m, $RMS<3$~ns and $N_{ex}<10$ resulting in 14267 imaged sources. The top panel shows height v.s.\ time of the sources where we have applied an off-set to put initiation close to $t=0$. The same sources, with the same coloring, are shown in the other panels giving height and distances north and east from the LOFAR core.}
	\figlab{2018-full}
\end{figure}

The complete 2018 flash, shown in \figref{2018-full}, is a typical example of a flash imaged with our techniques. The lightning flash (D20180813T153001) occurred on August 13, 2018 at 15:30 at a distance of about 50~km from the core of LOFAR, see \figref{LOFAR-NL}. To obtain this image we have used antennas for both polarizations and set the limits on the source quality as $\sigma(h)<3.5$~m, $RMS<3$~ns, and $N_{ex}<10$ from a total of about 265 antennas. This leaves 14267 imaged sources over the whole duration of the flash of 0.3~s. To give some idea of the effect of these limits we have relaxed the limit on the $RMS$ to $RMS<4$~ns which yields an image with an estimated 100 sources that are mislocated by about 100~m out of a total of 23606. Since the first imaged pulse of a flash is dependent on the applied source quality criteria we have not performed any fine-tuning in determining the time offset for \figref{2018-full}.

The flash shown in \figref{2018-full}, has a typical structure of flashes we have seen over the LOFAR area. The flash initiates at an altitude of about 5~km at the bottom part of the negative charge layer to develop first a Primary Initial Leader that triggers a number of negative leaders in the lower lying positive charge cloud. Only after about 20~ms the positive leader becomes visible in the form of increasing twinkle activity as we have reported in \cite{Hare:2019} for a different flash. The Primary Initial Leader forms a plasma channel that continues to serve as the link between the upper negative and lower positive charge layers and called the ``neck" is this work. The Initial Leader we observe is reminiscent of what is reported on in \cite<e.g.>{Lyu:2016} to occur at the initial phase of lightning storms in the US with the main difference that there the positive charge layer is positioned above the negative one.

\subsubsection{The initial development}\seclab{2018-Dyn}

\begin{figure*}[ht]
\centerline{\includegraphics[width=0.99\textwidth]{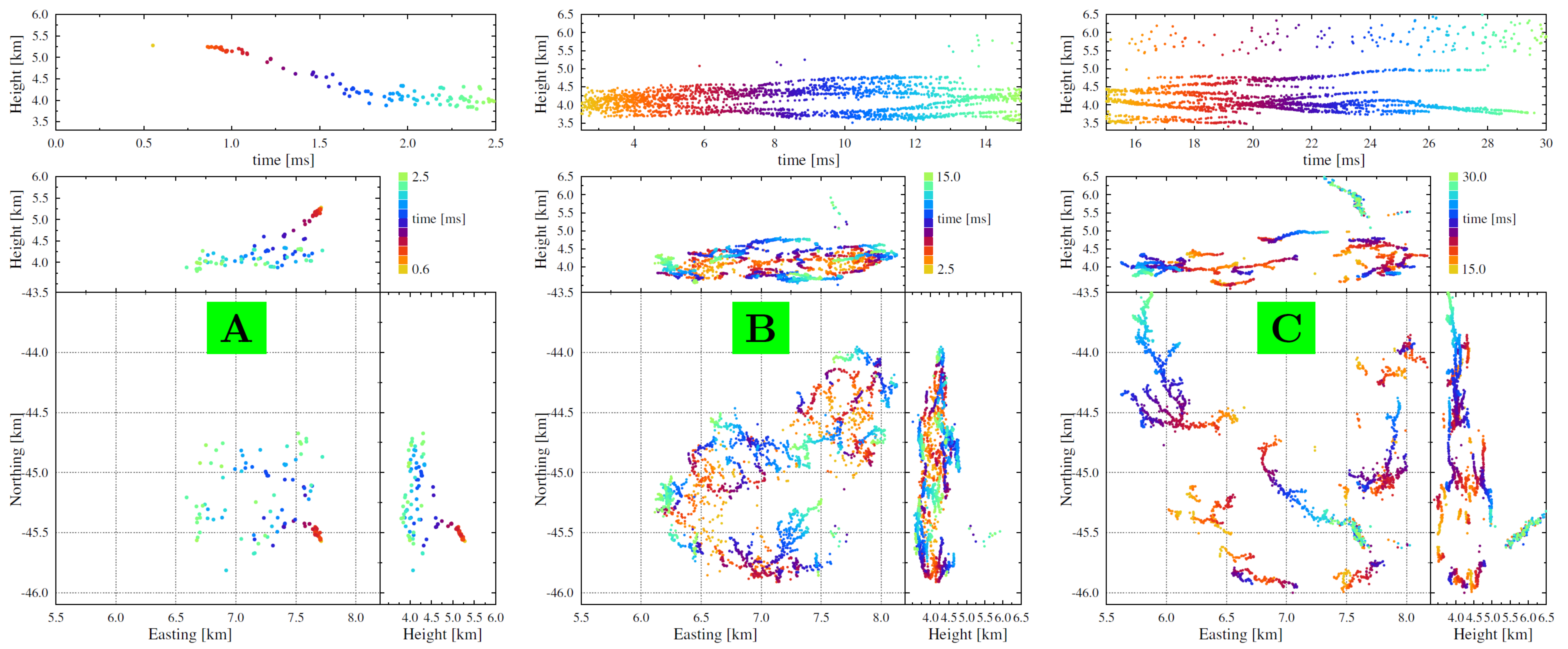} }
	\caption{The early part of the 2018 flash is imaged for three subsequent time periods to emphasize the dynamics in the initial stage.  Shown are sources with $\sigma(h)<3.5$~m, $RMS<4$~ns and $N_{ex}<20$ resulting in 113, 2849, 2814 imaged sources respectively for the three different time frames. After initiation at an height of 5.2~km one observes a fast downward progression (A). At an height of about 4~km, the downward motion stops and negative leaders develop in seemingly arbitrary directions at multiple places over an area in excess of 1~km$^2$ (B). At the end only one continues to grow (C). 3 time frames, not panels; figures need some work, we could show also subsequent time frames.}
	\figlab{2018-Init0-3s}
\end{figure*}

To visualize the dynamics of the leader development after initiation, \figref{2018-Init0-3s} shows the time development in chronological frames. Time frame A shows the Primary Initial Leader starting at an initial height of 5.25~km developing downward. For the first 0.5~ms no progression is observed but then it accelerates and progresses downward along a slanted path at about $1.2\times 10^6$~m/s, somewhat faster than observed in \cite{Lyu:2016}. After a kink in the path at 5~km altitude, (visible most clearly in the north v.s.\ altitude plot) the Primary Initial Leader fans out, initiating a multitude of negative leaders at an altitude of about 4.5 -- 4.0~km where the downward motion stops (see time frames B and C). Then distinct, almost horizontal, leaders develop in the same fashion as negative leaders do with a speed of about $10^5$~m/s, i.e.\ ten times slower than the initial leader. It is impressive that the formation of negative leaders starts simultaneously at multiple places over an area in excess of 1~km$^2$ (B). Each of the new leaders appears to develop in seemingly arbitrary directions, some inward, some outward. Close inspection shows that they rather seem to cover the surface of a spatial structure. One also notices that there is some VHF activity along the path of the initial leader at 6 and 8~ms and 5.2 -- 5.3~km height. Since this particular region in space is observed to play a rather special role in the evolution of the flash we named it the neck. The observed VHF emission indicates that current is flowing through the neck, even though there is no visible activity of a positive leader yet. Only after 13~ms the positive leader starts to show at an angle w.r.t.\ the Primary Initial Leader. The neck will serve as the connection point between the upper negative and the lower positive charge for the whole duration of the flash.
Time frame C shows a pronounced positive leader with ample twinkling activity along several needles \cite{Hare:2019,Pu:2019}. The neck itself shows no needle activity. It is interesting to see that one negative leader propagated to within 500~m horizontal distance and at the same altitude of the place where the neck showed a kink and started to fan out. Eventually there is only a single negative leader that continues to spread away from the initiation point, all others appear to have stalled.

\subsubsection{The Primary Initial Leader}\seclab{2018-PIL}

\begin{figure}[h]
\centerline{\includegraphics[width=0.79\textwidth]{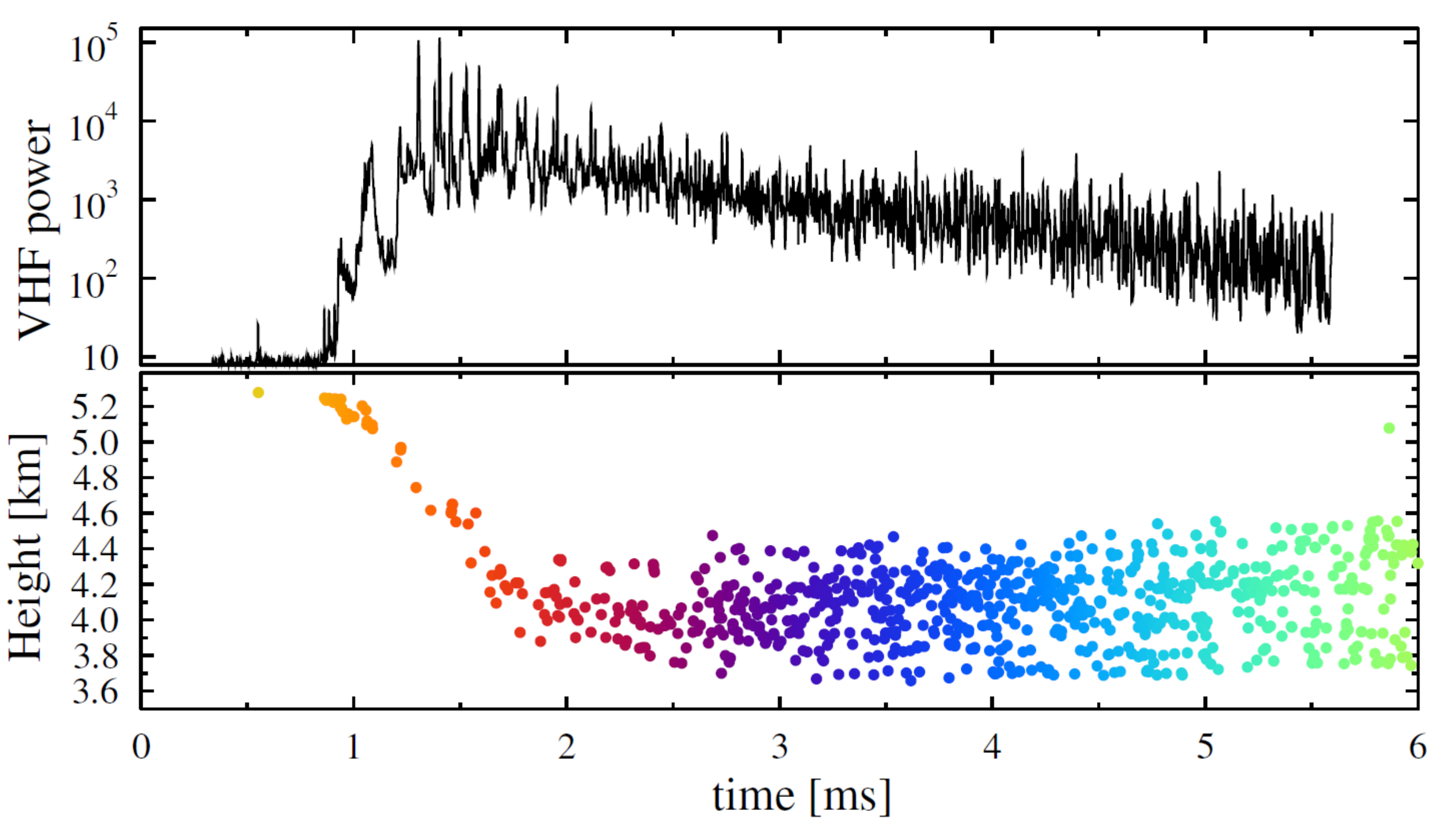} }
	\caption{The VHF power in arbitrary units in 4~$\mu$s bins, is compared to imaged sources for the first 6~ms of the 2018 flash.}
	\figlab{H-env-Ed}
\end{figure}

\figref{H-env-Ed} shows the Primary Initial Leader with the located sources together and the recorded power. 
The power is calculated as the square of the measured VHF-signal (the same as used in imaging) from on one of LOFAR's core antennas and averaged over 4~$\mu$s.
It is interesting to see that the VHF activity increases rapidly after initiation, reaching a peak at the time the Initial Leader creates negative leaders. This takes place 2~ms after initiation. At later times the VHF activity decreases again.

\begin{figure}[h]
\centerline{\includegraphics[width=0.69\textwidth]{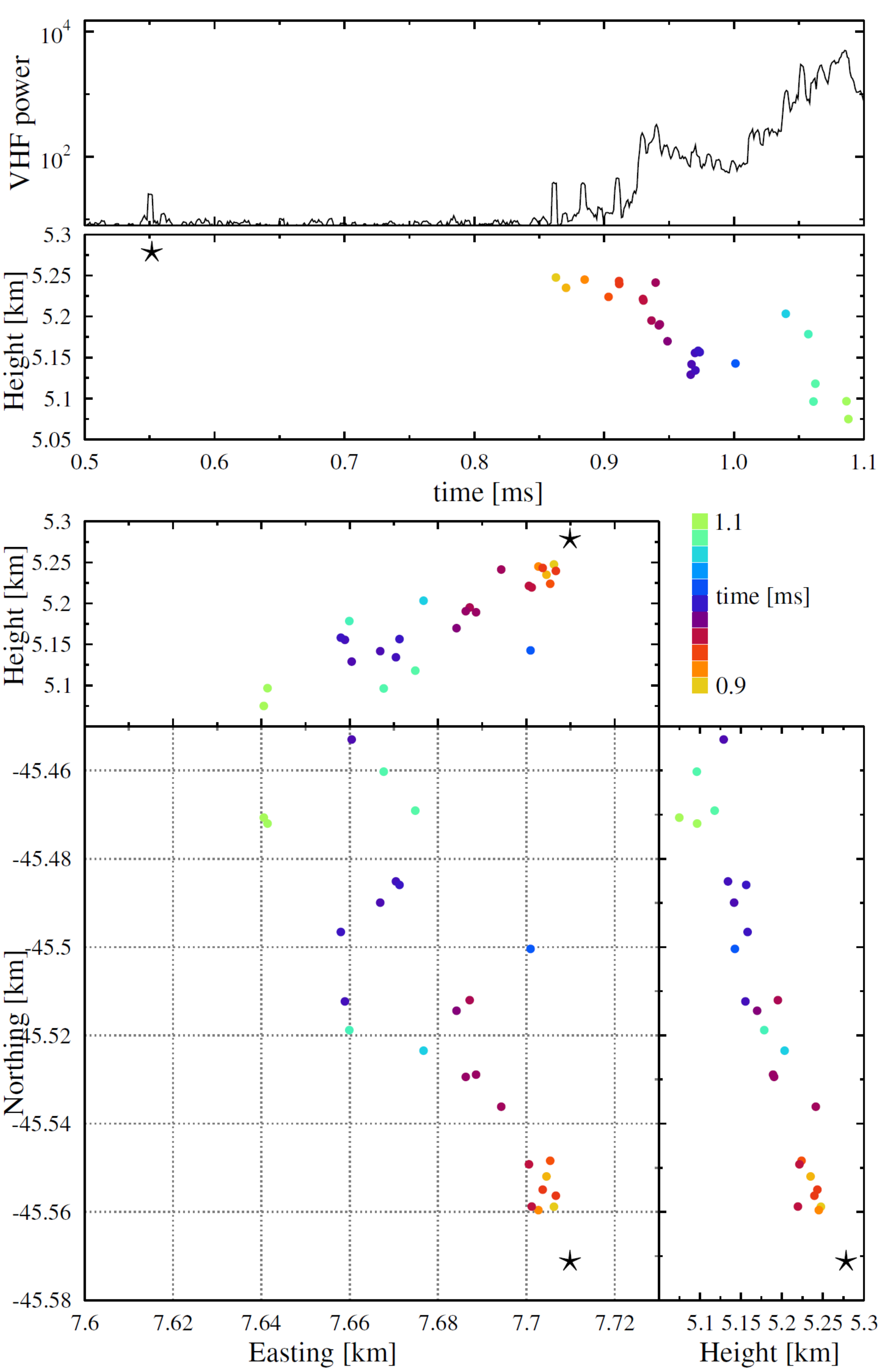} }
	\caption{A zoom-in on the sources observed in the initial 1.1~ms of the 2018 flash. The first located source, the initiation point, is marked with a black star.}
	\figlab{2018-IniZ}
\end{figure}

Making use of our high resolution, \figref{2018-IniZ} shows a zoom in of the initial leader to provide a better look at its properties.
From the very first source, marked with a star, the initial leader appears to fan out reaching a diameter of about 100~m in height and less in horizontal directions.
In the first part of the Initial Leader development a few different stages can be distinguished. The first stage ranges from initiation, $t=0.55$~ms, till $t=0.94$~ms. The VHF trace shows almost no power above background. The imager finds several good quality sources towards the end of this period. At this first stage the leader moves over a distance of 20~m horizontally and 50~m downward at a speed of $1.3\times 10^5$~m/s.
In the subsequent second stage, lasting from $t=0.94$ -- 0.97~ms an increased VHF activity is visible. During this stage the leader moves over a distance of 80~m horizontally and 90~m downward with a speed of $4\times 10^6$~m/s, considerably faster than in the first stage.
The third phase lasts till about 1.1~ms where one sees a clear first burst of VHF intensity.
The few sources (6) that are located at this stage are lying on a continuation of the leader seen at the second stage or around the previous leader.
In the fourth stage from $t=1.1$ -- 1.8~ms the Initial Leader continues to propagate down and continues to fan-out. The few sources we image in the interval from 1.5 -- 1.8~ms are spread over a slanted disk with a size of about 300~m in north-south as well as east-west. At this stage the VHF emission reaches a maximum and the high density of pulses prevents us from performing efficient mapping. In this stage the speed is about $2\times 10^6$~m/s.
At later stages the general downward motion stops, the propagation speed decreases, the VHF intensity continues to drop, and we are able to map an increasing number of sources. From the located sources we observe that a multitude of negative leaders branch off from the PIL, and then propagate at the typical speed of a negative leader, $10^5$~m/s.

\subsubsection{Initiation pulse}\seclab{2018-pulse}

\begin{figure}[h]
	\centering	\includegraphics[width=0.79\textwidth]{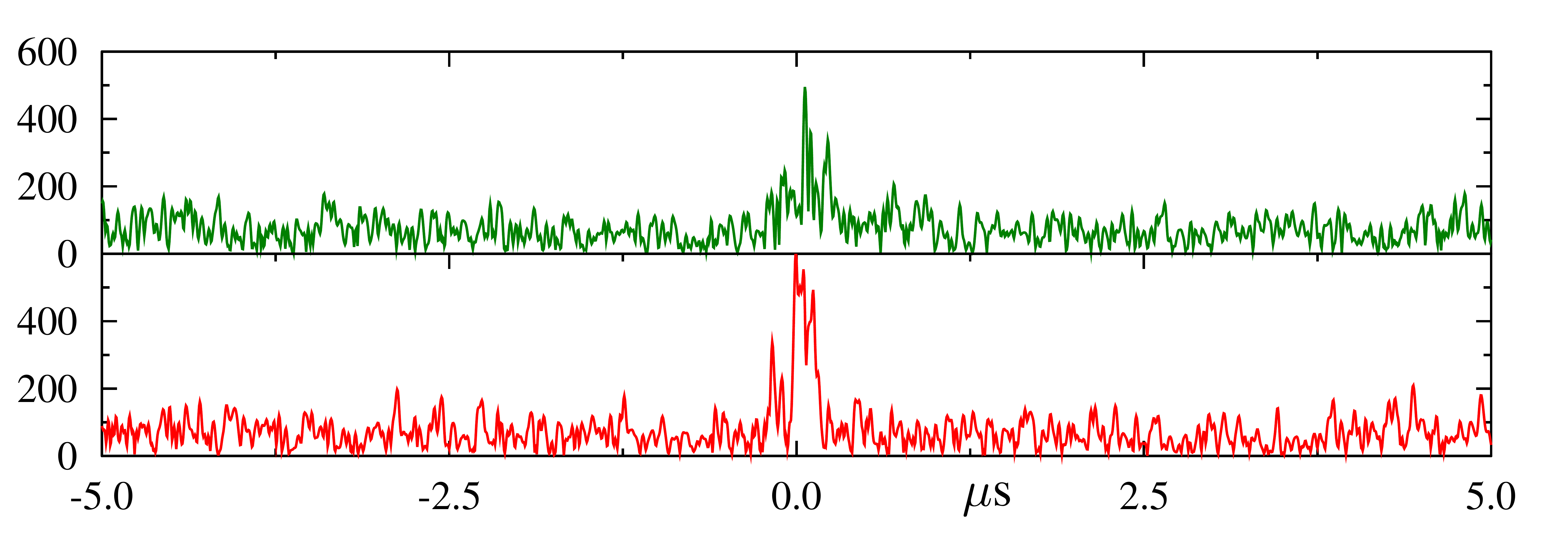}
	\caption{The Hilbert-envelope of the trace for each of the two polarization directions as measured at the core of LOFAR. The vertical axis gives pulse amplitude in arbitrary units, the horizontal axis time (in $\mu$s) centered at the first located source (at 0.55~ms in \figref{2018-IniZ}).}
	\figlab{pulse}
\end{figure}

To show that the flash activity already started at 0.55~ms we show in \figref{pulse} a part of time trace in the vicinity of this first imaged source. Here the pulse stands out clearly and, because the spectrum is relatively clean, we can observe it in all antennas. Also some even smaller pulses can be imaged, but these do not pass the quality criteria used in making \figref{2018-Init0-3s} and \figref{2018-IniZ}.
It should be noted that the trace in the two polarization directions is rather different, signalling that this first imaged pulse is due to a complicated current pattern, with currents in multiple directions. In addition this pulse is considerably longer than the pulse response of our system (which is 50~ns FWHM) which is additional evidence that its source is composite, not just a single short pulse.
We observe that the sub-structure of the pulse does not change significantly for antennas at different orientations w.r.t.\ the source (taking into account the polarization). Based on the pulse response of the system of 50~ns (15~m length at the speed of light) we thus conclude that the spatial extent of the source must be small, of the order or less than $(10 {\rm m})^3$.

\subsection{The 2019 flash}\seclab{2019}

\begin{figure}[h]
	\centering	\includegraphics[bb=0.5cm 2cm 22.5cm 22.5cm,clip, width=0.99\textwidth]{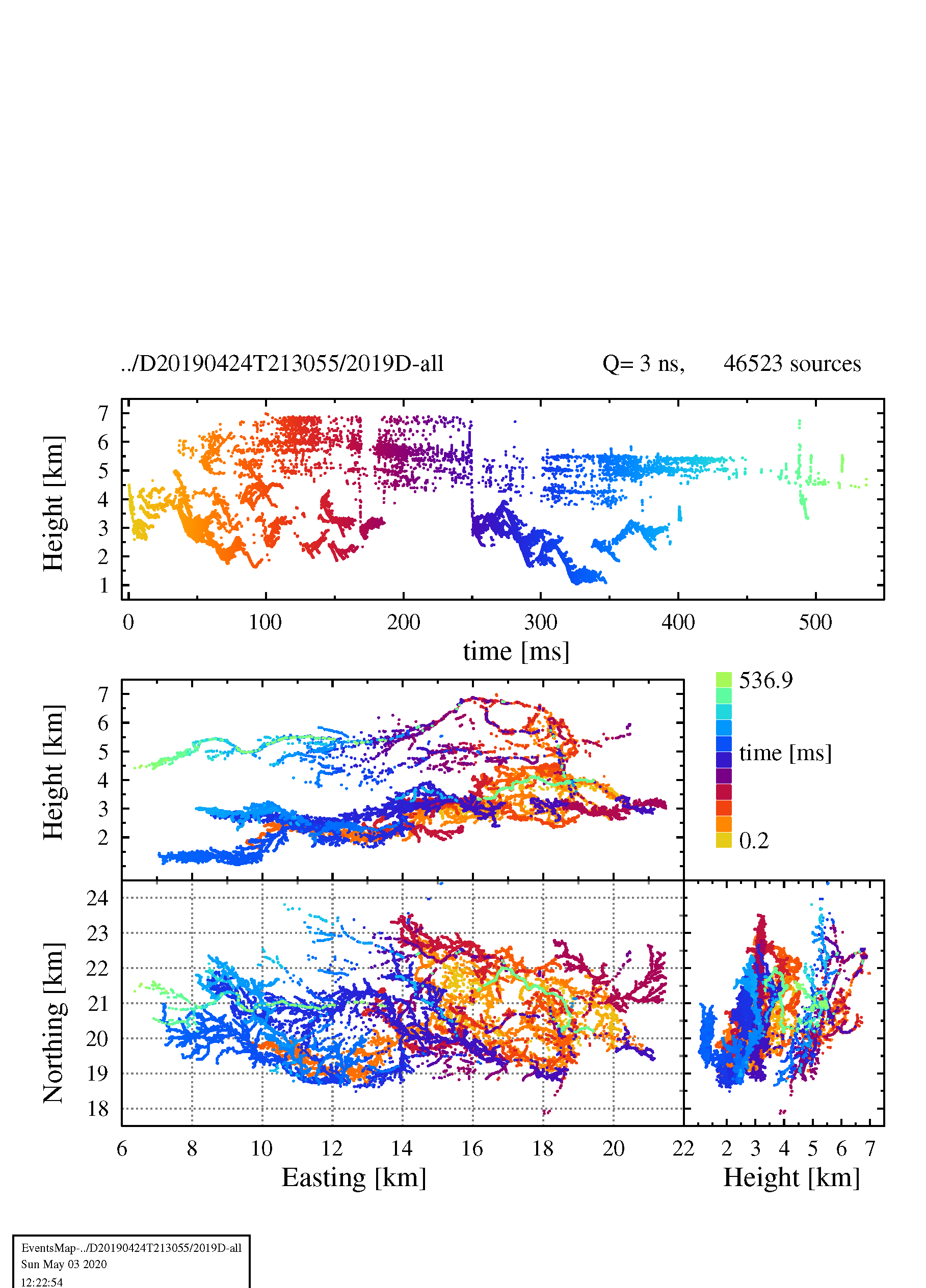}
	\caption{Image of the 2019 flash showing sources  with $\sigma(h)<3.5$~m, $RMS<3$~ns and $N_{ex}<25$ (out of a total of close to 400) are shown, resulting in 46523 imaged sources. }
	\figlab{2019-all}
\end{figure}

The complete image of the 2019 flash we discuss in this paper, D20190424T213055, occurring on April 24, 2019 at 21:30 is shown in \figref{2019-all}.
The time is shifted such that the flash starts close to $t=0$. Likely due to the close proximity to the core of LOFAR, we could image for this flash a larger density of sources per ms of the flash.
The flash shows the same charge-layer structure as was seen for the 2018 flash discussed in the previous section. For this flash the positive charge layer extends from about 4~km almost to the ground.
The negative charge layer does not extend above 7~km height.
The currents from the lower positive to the upper negative layer all appear to flow through the neck that was formed at initiation. In \secref{2019-Neck} we will show that within 100~m from the neck we observed a negative leader propagating to the positive charge layer.

\subsubsection{Secondary Initial Leaders}\seclab{2019-SIL}

\begin{figure*}[th]
\centerline{\includegraphics[width=0.99\textwidth]{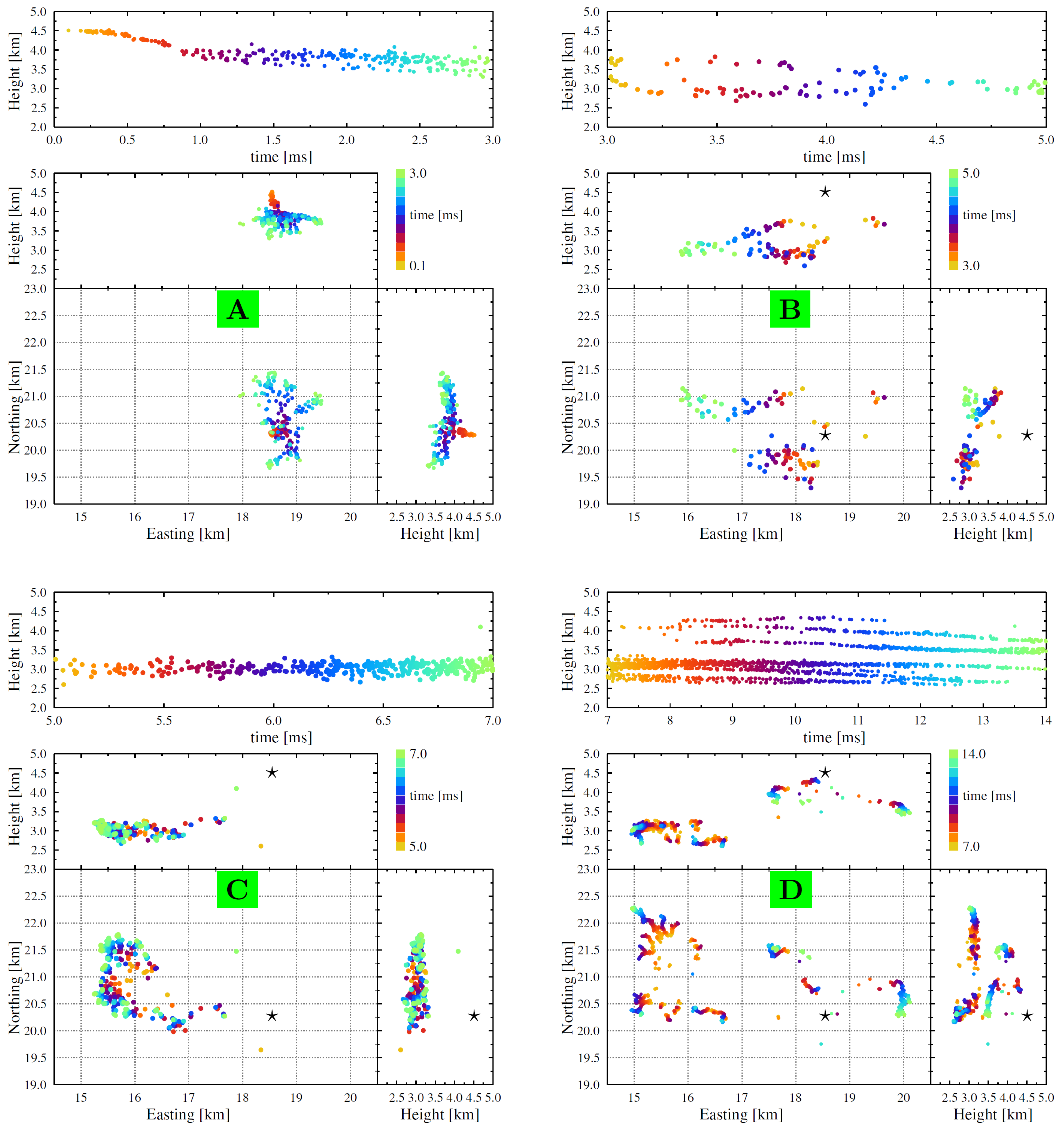} }
	\caption{The initiation of the 2019 flash.  Setting the image quality at $\sigma(h)<3.5$~m, $RMS<5$~ns and $N_{ex}<35$ leaves 323,99,413, and 2332 sources for the different sequential time frames. The black star in time frames B -- D marks the position of the point of initiation.   
}
	\figlab{2019-Init0-3s}
\end{figure*}

The initiation phase of the 2019 event is shown in detail in \figref{2019-Init0-3s}. 
It has been verified that there is no distinct peak visible in the time traces before the time of the first imaged source at $t=0.1$~ms. For \figref{2019-Init0-3s}, we have relaxed the condition on the source quality a little to increase the number of imaged sources, and made sure that the cuts were such that there were no obviously mislocated sources.
Time frame A in \figref{2019-Init0-3s} shows that after initiation the Primary Initial Leader propagates downward in an accelerated motion reaching a speed of about $2\times 10^6$~m/s rather straight down this time. At an altitude of about 4~km it starts to fan out, producing a multitude of negative leaders over an area of about 1~km$^2$. This is qualitatively the same as was observed for the 2018 event. Time frame B shows that after 3~ms most of the negative leaders stop propagating, while two of them show a fast motion covering a distance of 2~km in 2~ms, ten times faster than the propagation speed of the negative leaders ($10^5$~m/s) and the same as that of the Primary Initial Leader. Another resemblance is that the number of imaged sources on this leader is relatively low. For this reason we call them Secondary Initial Leaders. The bottom-left panel shows that when they reach a height of 3~km one of the Secondary Initial Leaders repeats the process of generating a multitude of negative leaders over an area in excess of 1~km$^2$, i.e.\ in the same positive charge layer situated a kilometer below and 2~km eastward than from the positive layer seen in time frame A.  The bottom-right panel shows that  most of the negative leaders in the second phase have stopped propagating, and that some of the initial group are re-activated.

\begin{figure}[h]
\centering{\includegraphics[bb=1.3cm 0.5cm 22.5cm 19cm,clip, width=0.79\textwidth]{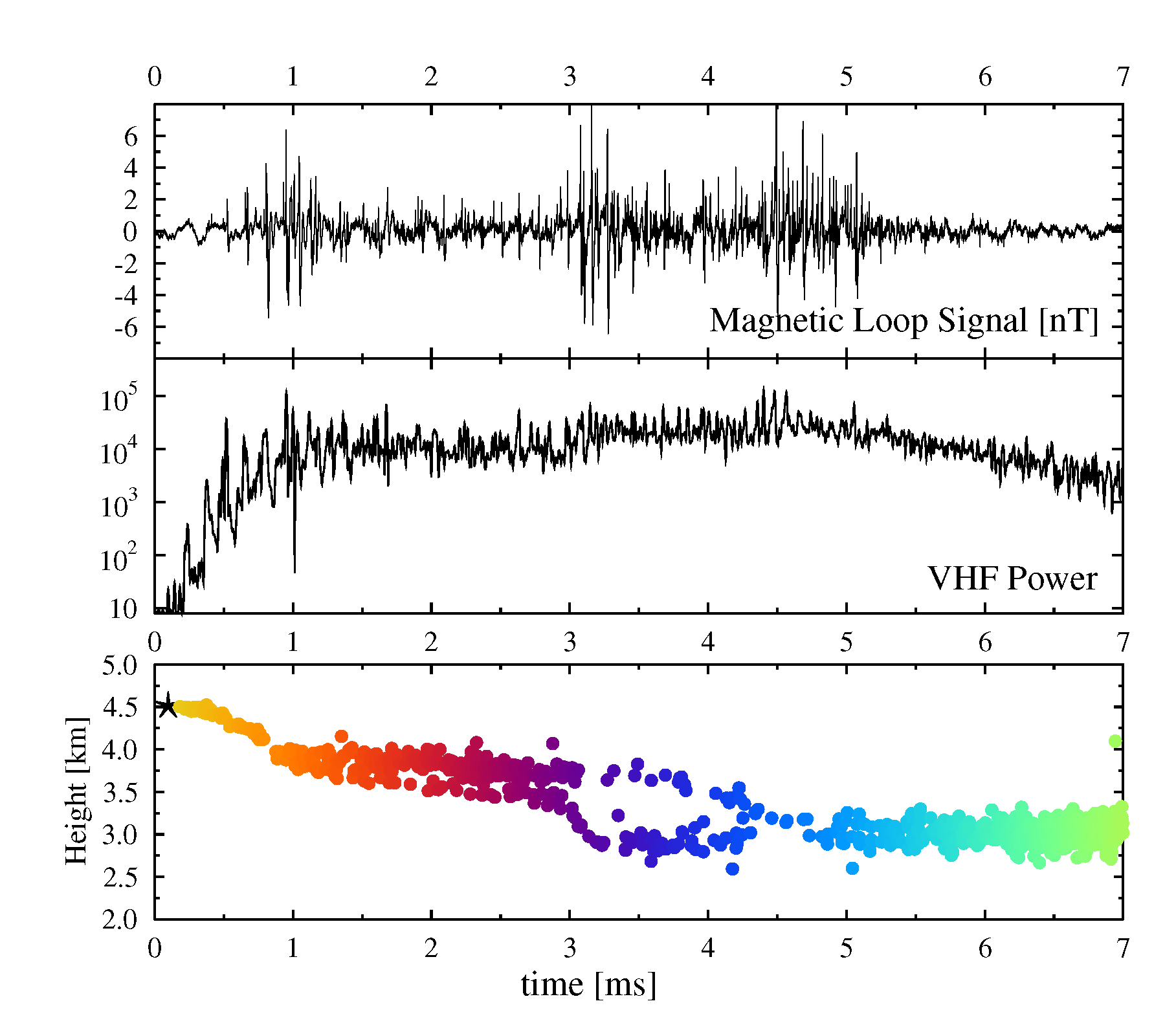} }
	\caption{The power recorded at a core station of LOFAR (re-binned over 2~$\mu$s, middle panel) is compared to the recording in the magnetic loop antenna (top panel, units of [nT]) and the height of the imaged sources shown for the first 7~ms in \figref{2019-Init0-3s}. All panels have been aligned in time. 
}
	\figlab{LoopAnt}
\end{figure}

At the time of initiation of this flash the magnetic loop antenna recorded data. The magnetic loop antenna is situated at a site some 10~km east from the LOFAR core, see \figref{LOFAR-NL}.
\figref{LoopAnt} shows the recorded magnetic loop antenna spectrum aligned with the imaged sources and the VHF power (averaged over 2~$\mu$s) as recorded by LOFAR. The figure shows that during the evolution of the Primary Initial Leader, in the first 1.5~ms, the VHF power increases in steps, identical to what was observed in \figref{2018-IniZ} for the 2018 flash. During this time the magnetic loop antenna recorded a dozen strong initial breakdown pulses.
The VHF power shows another strong increase (by a factor 2 or 3) at $t=3$~ms when the Secondary Initial Leaders start to emerge. Around the same time the magnetic loop antenna also measures an enhanced density of initial breakdown pulses.
There is another phase of enhanced activity seen in the ML antenna when the secondary initial leaders start to ignite the negative leaders around $t=4.5$~ms. At later times there is a gradual drop in the VHF power, at about the same rate as seen for the 2018 flash, of about one order of magnitude over 2~ms.

\figref{LoopAnt} also suggests that while the PIL is more gradually accelerated, the SIL moves immediately at the top speed, one straight down, the other initially horizontal.

\begin{figure}[h]
\centering{\includegraphics[bb=1.3cm 0.5cm 22.5cm 19cm,clip, width=0.79\textwidth]{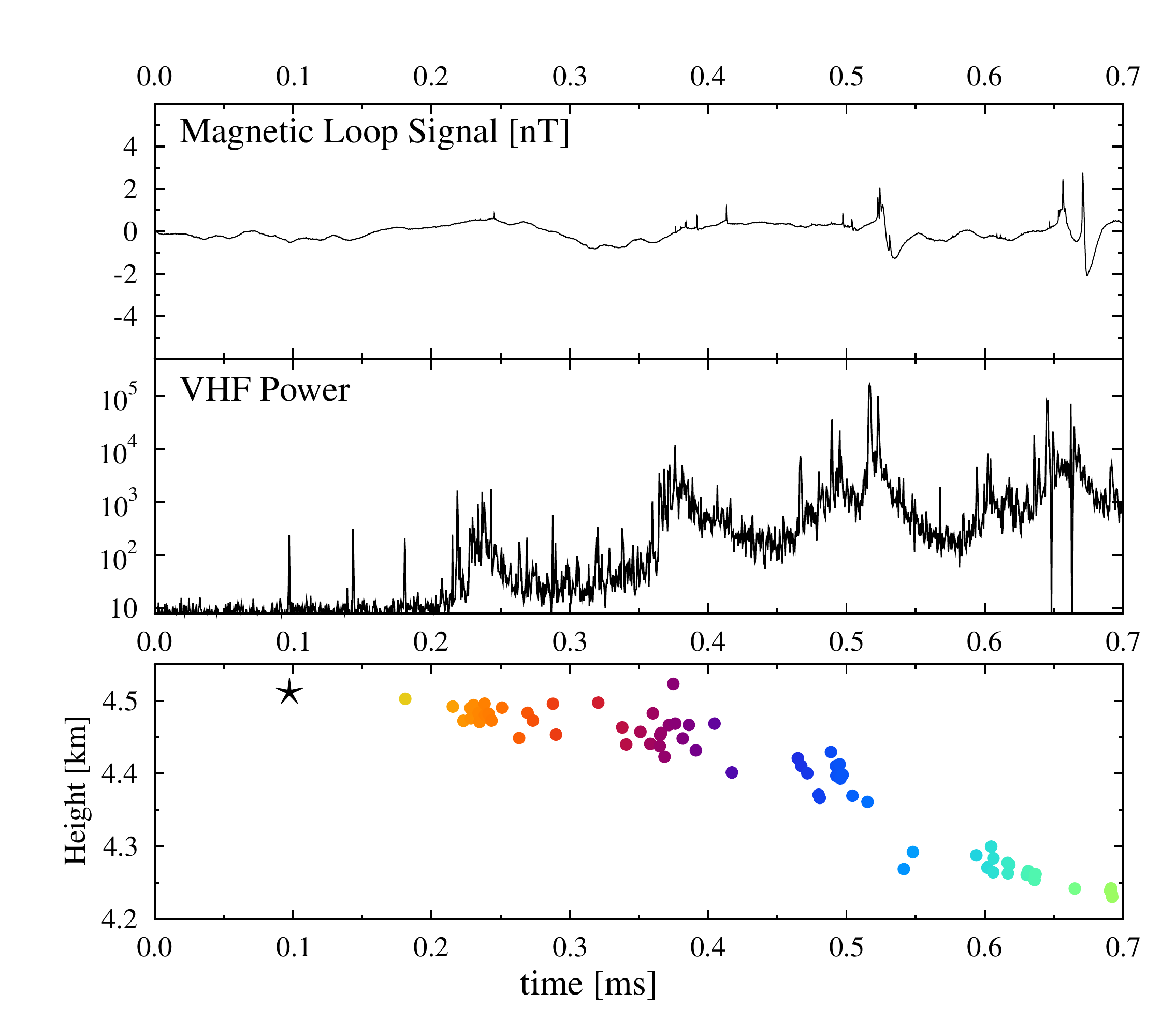} }
	\caption{The power recorded at a core station of LOFAR (re-binned over 0.5~$\mu$s, middle panel) is compared to the recording in the magnetic loop antenna (top panel) and the height of the imaged sources shown for the first 0.5~ms in \figref{2019-Init0-3s}. All panels have been aligned in time. 
}
	\figlab{LoopAnt05}
\end{figure}

Zooming in on the first millisecond flash \figref{LoopAnt05} shows many interesting aspects of the primary initial leader  of which we mention here only a few.

The broadband spectrum shows several pulses, each of which appears to be followed by a burst in VHF emission. This VHF emission is in the form of a enormous number of small pulses that combine in a strong increase in the emitted power. The full width at half maximum of such a burst, as can be determined from the first one, is about 0.01 -- 0.02~ms. This is very close to the burst duration seen in \cite{Hare:2020} where it is associated with negative leader stepping. However, the time between bursts of VHF power is about 0.1 -- 0.15~ms which is longer than the 0.05~ms observed in \cite{Hare:2020} for normal negative leader propagation.

From \figref{2019-Init0-3s} it can be seen that the initial leader propagates almost vertically downward and the propagation speed can thus be deduced directly from \figref{LoopAnt05}. This indicates a constant acceleration of the PIL after it starts propagating, very similar to what was observed for the 2018 flash.

In a follow-up paper, \cite{Scholten:BBLofar}, the correspondence between the broad band signal and the LOFAR image will be discussed in more detail. There we will address beside the pulses seen in the broad band spectrum close to initiation also the pulses seen later during the flash.

\subsubsection{The Neck region}\seclab{2019-Neck}

\begin{figure}[h]
\centering{\includegraphics[ width=0.69\textwidth]{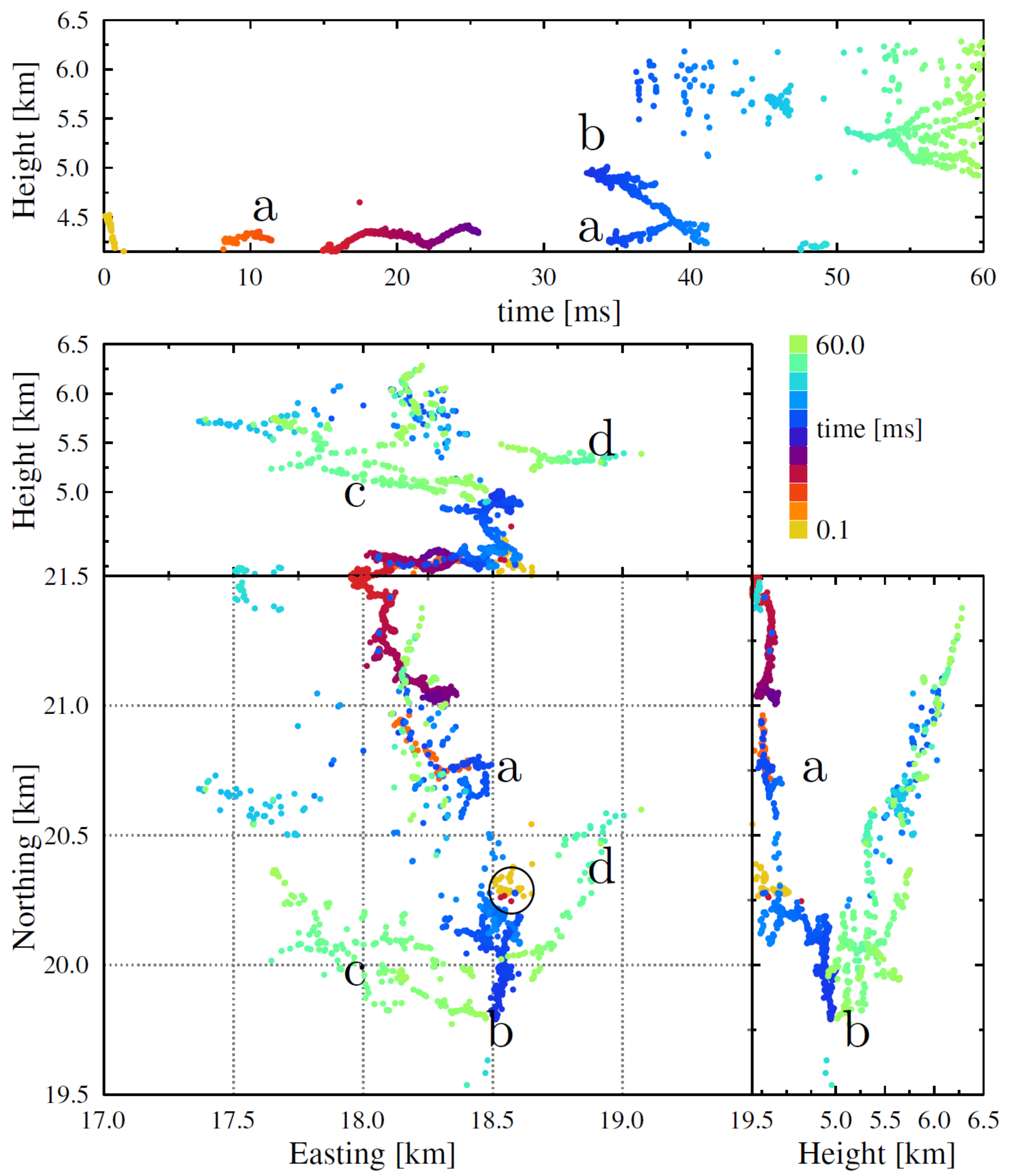} }
	\caption{An expanded view of the neck region for the 2019 flash. The letters are used in the text to refer to particular parts of this image. The initiation point is labeled by the $\bigcirc$.}
	\figlab{2019-neck}
\end{figure}

In the vicinity of the neck region, we have observed some interesting leader propagation that should be taken into account when reconstructing the structure of the charge layers in the vicinity of the initiation point labeled with a $\bigcirc$ in \figref{2019-neck}. At about $t=10$~ms a negative leader propagates up to the point (a) in the figure, a mere 500~m away from the initiation point horizontally. At $t=35$~ms the same negative leader activates a second time and approaches even to a distance of 300~m. At about the same time from point (b) at an altitude of 5~km a negative leader starts to propagate down, with no previous activity seen at its initiation point, towards the neck, to turn parallel to it at a distance of a mere 100~m where it stops propagating at the top of the charge layer where the first negative leaders were observed. At about the same time some VHF sources are observed along the positive leader. At $t=55$~ms at (c) a bit higher than 5~km another negative leader starts propagating towards point (b) where it appears to connect to the channel of the earlier negative leader that started from here. This very complicated leader structure seems to imply that around the initiation point a fair amount of charge was present in a complicated charge structure. In the 2018 flash we have also observed that a negative leader propagated up from the lower positive charge layer to the vicinity of the neck, but the other features appear to be unique for the 2019 flash.


\section{Discussion and conclusions}\seclab{discuss}

We have improved the procedure used in \cite{Hare:2019} for imaging lightning flashes using data from the LOFAR radio telescope. Our new imaging procedure has a proven capability to locate over 200 sources per millisecond of flash with  meter-scale accuracy.

We investigated the fist few ms of some flashes over the Netherlands and observe interesting leader structures, for example pointing to dense and relatively small charge clouds (see presentation).

All images of flashes we have made with LOFAR show a negative charge layer around 5~km and below this a positive charge layer which is consistent with the atmospheric electric fields over the LOFAR core as were determined in \cite{Trinh:2020}. We also observe an extensive structure of negative leaders in the lower positive charge layer. Very rarely do we see negative leader activity propagating up into the upper positive layer. This general structure is opposite to what is generally observed in US thunderstorms. 

In all our imaged flashes we observe a very similar initial development of the discharge where after initiation the discharge slowly grows (in fractions of a millisecond), picks up intensity (in the form of strong increase in emitted VHF power) as is typical for an initial leader. When the initial leader reaches the charge cloud which attracted it, the fast forward propagation stops and instead we observe the creation of many more normal negative leaders that propagate at about one-tenth of the speed of the initial leader.

This naturally leads to the picture that the initiation at the bottom side of a negative charge layer was driven by a relatively small (order 1~km$^3$) pocket of rather dense positive charge. This created the strong electric field that is needed to cause lightning initiation. In this strong field the elongating leader rapidly acquires charge by induction leading to increased currents and heating which could explain the initial acceleration we observe in the PIL. An acceleration of the initial leader was also observed in \cite{Cummer:2015}, although the acceleration was much smaller (about a factor 2 in velocity) than what we observe (close to an order of magnitude).
Since the ambient electric field extends up to the positive charge pocket, the fast propagation of the PIL stops as it reaches it and we start to observe the usual negative leaders that propagate along potential extremals \cite{Coleman:2008} and thus in a much weaker ambient field.

Our observations suggests that a relatively small positive charge pocket may be the driver for the strong electric field in which the initiation of the lightning happens. The presence of large hydro meteors, see \cite<e.g.>{Dubinova:2015}, or a large number of small droplets, see \cite{Kostinskiy:2019}, will be required as initiation sites. The initiation process is greatly facilitated by a high density of free electrons as created by cosmic rays as suggested in \cite{Rutjes:2019}.
It is not clear if any of these mechanisms can explain the initial accelerating propagation of the initial leader that we have observed.

The large number of negative leaders in a very confined space, as initiated by the PIL, are a strong indication that there must have been a pocket with a rather high charge density.
Once the charge in this confined charge cloud is collected the majority of the leaders stop propagating and only few remain. For the 2019 flash there may have been another positive charge cloud in the vicinity that attracts a secondary Initial Leader. As this is initiated from the tip of a propagating negative leader there is sufficient charge available to immediately have a fast-propagating discharge.

It has been suggested that turbulence is a very efficient mechanism in especially thunderclouds for creating regions with large fields that even grow exponentially with time \cite{Mareev:2017}. This could thus be the mechanism that created the relatively small pocket with high charge density.

Another indication of a dense and complicated charge layer structure in the vicinity of the initiation point is the fact that in our images we see a negative leader turning back towards the Neck, the initiation point of the flash. This happened for the 2018 as well as for the 2019 flash.

We observe a strong correlation between the signal of the broadband antenna and the emitted VHF power, as was already suggested in \cite{Kolmasova:2018,Kolmasova:2019}.
The relation between emitted VHF power, the signal of a broadband antenna and interferometry images  for the initial leader has been investigated in \cite{Krehbiel:2017} for New Mexico lightning flashes.


\acknowledgments

The LOFAR cosmic ray key science project acknowledges funding from an Advanced Grant of the European Research Council (FP/2007-2013) / ERC Grant Agreement n. 227610. 
The project has also received funding from the European Research Council (ERC) under the European Union's Horizon 2020 research and innovation programme (grant agreement No 640130). 
We furthermore acknowledge financial support from FOM, (FOM-project 12PR304). 
BMH is supported by the NWO (VI.VENI.192.071).
AN is supported by the DFG (NE 2031/2-1).
TW is supported by DFG grant 4946/1-1.
The work of IK and O.Santol\'{i}k was supported by European Regional Development Fund-Project CRREAT (CZ.02.1.01/0.0/0.0/15-003/0000481) and by the Praemium Academiae award of the Czech Academy of Sciences AS.
The work of  RL and LU was supported by the GACR grant 20-09671S.
KM is supported by FWO (FWOTM944). 
TNGT acknowledges funding from the Vietnam National Foundation for Science and Technology Development (NAFOSTED) under Grant 103.01-2019.378.
ST acknowledges funding from the Khalifa University Startup grant (project code 8474000237).
\\
This paper is based on data obtained with the International LOFAR Telescope (ILT). LOFAR~\cite{Haarlem:2013} is the Low Frequency Array designed and constructed by ASTRON. It has observing, data processing, and data storage facilities in several countries, that are owned by various parties (each with their own funding sources), and that are collectively operated by the ILT foundation under a joint scientific policy. The ILT resources have benefitted from the following recent major funding sources: CNRS-INSU, Observatoire de Paris and Universit\'{e} d'Orl\'{e}ans, France; BMBF, MIWF-NRW, MPG, Germany; Science Foundation Ireland (SFI), Department of Business, Enterprise and Innovation (DBEI), Ireland; NWO, The Netherlands; The Science and Technology Facilities Council, UK.

The data are available from the LOFAR Long Term Archive (lta.lofar.eu), under the following locations:
\\ \noindent \verb!L664246_D20180813T153001.413Z_stat_R000_tbb.h5!
\\ \noindent \verb!L703974_D20190424T213055.202Z_stat_R000_tbb.h5!
\\ \noindent all of them with the same prefix
\\ \noindent \verb!srm.grid.sara.nl/pnfs/grid.sara.nl/data/lofar/ops/TBB/lightning/! and where  ``stat'' should be replaced by the name of the station, CS001, CS002, CS003, CS004, CS005, CS006, CS007, CS011, CS013, CS017, CS021, CS024, CS026, CS030, CS031, CS032, CS101, CS103, RS106, CS201, RS205, RS208, RS210, CS301, CS302, RS305, RS306, RS307, RS310, CS401, RS406, RS407, RS409, CS501, RS503, or RS508.

\bibliography{LightInit-v7c}

\begin{thebibliography}{}

\bibitem [\protect \citeauthoryear {%
\APACcitebtitle {Blitzortung.org}}{%
\APACcitebtitle {Blitzortung.org}}{%
{\protect \APACyear {{\protect \bibnodate {}}}}%
}]{%
Blitzortung}
\APACinsertmetastar {%
Blitzortung}%
\APACrefbtitle {Blitzortung.org.} {Blitzortung.org.}
\newblock
\APACrefYearMonthDay{{\protect \bibnodate {}}}{}{}.
\newblock
\APAChowpublished {\url{http://LightningMaps.org}}.
\PrintBackRefs{\CurrentBib}

\bibitem [\protect \citeauthoryear {%
Coleman%
, Stolzenburg%
, Marshall%
\BCBL {}\ \BBA {} Stanley%
}{%
Coleman%
\ \protect \BOthers {.}}{%
{\protect \APACyear {2008}}%
}]{%
Coleman:2008}
\APACinsertmetastar {%
Coleman:2008}%
\begin{APACrefauthors}%
Coleman, L\BPBI M.%
, Stolzenburg, M.%
, Marshall, T\BPBI C.%
\BCBL {}\ \BBA {} Stanley, M.%
\end{APACrefauthors}%
\unskip\
\newblock
\APACrefYearMonthDay{2008}{}{}.
\newblock
{\BBOQ}\APACrefatitle {Horizontal lightning propagation, preliminary breakdown,
  and electric potential in New Mexico thunderstorms} {Horizontal lightning
  propagation, preliminary breakdown, and electric potential in new mexico
  thunderstorms}.{\BBCQ}
\newblock
\APACjournalVolNumPages{Journal of Geophysical Research:
  Atmospheres}{113}{D9}{}.
\newblock
\begin{APACrefURL}
  \url{https://agupubs.onlinelibrary.wiley.com/doi/abs/10.1029/2007JD009459}
  \end{APACrefURL}
\newblock
\begin{APACrefDOI} \doi{10.1029/2007JD009459} \end{APACrefDOI}
\PrintBackRefs{\CurrentBib}

\bibitem [\protect \citeauthoryear {%
Cummer%
\ \protect \BOthers {.}}{%
Cummer%
\ \protect \BOthers {.}}{%
{\protect \APACyear {2015}}%
}]{%
Cummer:2015}
\APACinsertmetastar {%
Cummer:2015}%
\begin{APACrefauthors}%
Cummer, S\BPBI A.%
, Lyu, F.%
, Briggs, M\BPBI S.%
, Fitzpatrick, G.%
, Roberts, O\BPBI J.%
\BCBL {}\ \BBA {} Dwyer, J\BPBI R.%
\end{APACrefauthors}%
\unskip\
\newblock
\APACrefYearMonthDay{2015}{}{}.
\newblock
{\BBOQ}\APACrefatitle {Lightning leader altitude progression in terrestrial
  gamma-ray flashes} {Lightning leader altitude progression in terrestrial
  gamma-ray flashes}.{\BBCQ}
\newblock
\APACjournalVolNumPages{Geophysical Research Letters}{42}{18}{7792-7798}.
\newblock
\begin{APACrefURL}
  \url{https://agupubs.onlinelibrary.wiley.com/doi/abs/10.1002/2015GL065228}
  \end{APACrefURL}
\newblock
\begin{APACrefDOI} \doi{10.1002/2015GL065228} \end{APACrefDOI}
\PrintBackRefs{\CurrentBib}

\bibitem [\protect \citeauthoryear {%
Dubinova%
\ \protect \BOthers {.}}{%
Dubinova%
\ \protect \BOthers {.}}{%
{\protect \APACyear {2015}}%
}]{%
Dubinova:2015}
\APACinsertmetastar {%
Dubinova:2015}%
\begin{APACrefauthors}%
Dubinova, A.%
, Rutjes, C.%
, Ebert, U.%
, Buitink, S.%
, Scholten, O.%
\BCBL {}\ \BBA {} Trinh, G\BPBI T\BPBI N.%
\end{APACrefauthors}%
\unskip\
\newblock
\APACrefYearMonthDay{2015}{Jun}{}.
\newblock
{\BBOQ}\APACrefatitle {{Prediction of Lightning Inception by Large Ice
  Particles and Extensive Air Showers}} {{Prediction of Lightning Inception by
  Large Ice Particles and Extensive Air Showers}}.{\BBCQ}
\newblock
\APACjournalVolNumPages{Phys. Rev. Lett.}{115}{}{015002}.
\newblock
\begin{APACrefDOI} \doi{10.1103/PhysRevLett.115.015002} \end{APACrefDOI}
\PrintBackRefs{\CurrentBib}

\bibitem [\protect \citeauthoryear {%
Edens%
\ \protect \BOthers {.}}{%
Edens%
\ \protect \BOthers {.}}{%
{\protect \APACyear {2012}}%
}]{%
Edens:2012}
\APACinsertmetastar {%
Edens:2012}%
\begin{APACrefauthors}%
Edens, H\BPBI E.%
, Eack, K\BPBI B.%
, Eastvedt, E\BPBI M.%
, Trueblood, J\BPBI J.%
, Winn, W\BPBI P.%
, Krehbiel, P\BPBI R.%
\BDBL {}Thomas, R\BPBI J.%
\end{APACrefauthors}%
\unskip\
\newblock
\APACrefYearMonthDay{2012}{}{}.
\newblock
{\BBOQ}\APACrefatitle {VHF lightning mapping observations of a triggered
  lightning flash} {Vhf lightning mapping observations of a triggered lightning
  flash}.{\BBCQ}
\newblock
\APACjournalVolNumPages{Geophysical Research Letters}{39}{19}{}.
\newblock
\begin{APACrefDOI} \doi{10.1029/2012GL053666} \end{APACrefDOI}
\PrintBackRefs{\CurrentBib}

\bibitem [\protect \citeauthoryear {%
B.~Hare%
\ \protect \BOthers {.}}{%
B.~Hare%
\ \protect \BOthers {.}}{%
{\protect \APACyear {2019}}%
}]{%
Hare:2019}
\APACinsertmetastar {%
Hare:2019}%
\begin{APACrefauthors}%
Hare, B.%
\BCBT {}\ \BOthersPeriod {.}
\end{APACrefauthors}%
\unskip\
\newblock
\APACrefYearMonthDay{2019}{}{}.
\newblock
{\BBOQ}\APACrefatitle {{Needle-like structures discovered on positively charged
  lightning branches}} {{Needle-like structures discovered on positively
  charged lightning branches}}.{\BBCQ}
\newblock
\APACjournalVolNumPages{Nature}{568}{}{360--363}.
\newblock
\begin{APACrefDOI} \doi{10.1038/s41586-019-1086-6} \end{APACrefDOI}
\PrintBackRefs{\CurrentBib}

\bibitem [\protect \citeauthoryear {%
B\BPBI M.~Hare%
\ \protect \BOthers {.}}{%
B\BPBI M.~Hare%
\ \protect \BOthers {.}}{%
{\protect \APACyear {2018}}%
}]{%
Hare:2018}
\APACinsertmetastar {%
Hare:2018}%
\begin{APACrefauthors}%
Hare, B\BPBI M.%
\BCBT {}\ \BOthersPeriod {.}
\end{APACrefauthors}%
\unskip\
\newblock
\APACrefYearMonthDay{2018}{}{}.
\newblock
{\BBOQ}\APACrefatitle {{LOFAR Lightning Imaging: Mapping Lightning With
  Nanosecond Precision}} {{LOFAR Lightning Imaging: Mapping Lightning With
  Nanosecond Precision}}.{\BBCQ}
\newblock
\APACjournalVolNumPages{Journal of Geophysical Research:
  Atmospheres}{123}{5}{2861-2876}.
\newblock
\begin{APACrefDOI} \doi{10.1002/2017JD028132} \end{APACrefDOI}
\PrintBackRefs{\CurrentBib}

\bibitem [\protect \citeauthoryear {%
B\BPBI M.~Hare%
\ \protect \BOthers {.}}{%
B\BPBI M.~Hare%
\ \protect \BOthers {.}}{%
{\protect \APACyear {2020}}%
}]{%
Hare:2020}
\APACinsertmetastar {%
Hare:2020}%
\begin{APACrefauthors}%
Hare, B\BPBI M.%
, Scholten, O.%
, Dwyer, J.%
, Ebert, U.%
, Nijdam, S.%
, Bonardi, A.%
\BDBL {}Winchen, T.%
\end{APACrefauthors}%
\unskip\
\newblock
\APACrefYearMonthDay{2020}{Mar}{}.
\newblock
{\BBOQ}\APACrefatitle {Radio Emission Reveals Inner Meter-Scale Structure of
  Negative Lightning Leader Steps} {Radio emission reveals inner meter-scale
  structure of negative lightning leader steps}.{\BBCQ}
\newblock
\APACjournalVolNumPages{Phys. Rev. Lett.}{124}{}{105101}.
\newblock
\begin{APACrefURL}
  \url{https://link.aps.org/doi/10.1103/PhysRevLett.124.105101}
  \end{APACrefURL}
\newblock
\begin{APACrefDOI} \doi{10.1103/PhysRevLett.124.105101} \end{APACrefDOI}
\PrintBackRefs{\CurrentBib}

\bibitem [\protect \citeauthoryear {%
Hill%
, Uman%
\BCBL {}\ \BBA {} Jordan%
}{%
Hill%
\ \protect \BOthers {.}}{%
{\protect \APACyear {2011}}%
}]{%
Hill:2011}
\APACinsertmetastar {%
Hill:2011}%
\begin{APACrefauthors}%
Hill, J\BPBI D.%
, Uman, M\BPBI A.%
\BCBL {}\ \BBA {} Jordan, D\BPBI M.%
\end{APACrefauthors}%
\unskip\
\newblock
\APACrefYearMonthDay{2011}{}{}.
\newblock
{\BBOQ}\APACrefatitle {High-speed video observations of a lightning stepped
  leader} {High-speed video observations of a lightning stepped leader}.{\BBCQ}
\newblock
\APACjournalVolNumPages{Journal of Geophysical Research:
  Atmospheres}{116}{D16}{}.
\newblock
\begin{APACrefDOI} \doi{10.1029/2011JD015818} \end{APACrefDOI}
\PrintBackRefs{\CurrentBib}

\bibitem [\protect \citeauthoryear {%
Kolma\v{s}ov\'{a}%
\ \protect \BOthers {.}}{%
Kolma\v{s}ov\'{a}%
\ \protect \BOthers {.}}{%
{\protect \APACyear {2019}}%
}]{%
Kolmasova:2019}
\APACinsertmetastar {%
Kolmasova:2019}%
\begin{APACrefauthors}%
Kolma\v{s}ov\'{a}, I.%
, Marshall, T.%
, Bandara, S.%
, Karunarathne, S.%
, Stolzenburg, M.%
, Karunarathne, N.%
\BCBL {}\ \BBA {} Siedlecki, R.%
\end{APACrefauthors}%
\unskip\
\newblock
\APACrefYearMonthDay{2019}{}{}.
\newblock
{\BBOQ}\APACrefatitle {Initial Breakdown Pulses Accompanied by VHF Pulses
  During Negative Cloud-to-Ground Lightning Flashes} {Initial breakdown pulses
  accompanied by vhf pulses during negative cloud-to-ground lightning
  flashes}.{\BBCQ}
\newblock
\APACjournalVolNumPages{Geophysical Research Letters}{46}{10}{5592-5600}.
\newblock
\begin{APACrefURL}
  \url{https://agupubs.onlinelibrary.wiley.com/doi/abs/10.1029/2019GL082488}
  \end{APACrefURL}
\newblock
\begin{APACrefDOI} \doi{10.1029/2019GL082488} \end{APACrefDOI}
\PrintBackRefs{\CurrentBib}

\bibitem [\protect \citeauthoryear {%
Kolma\v{s}ov\'{a}%
\ \protect \BOthers {.}}{%
Kolma\v{s}ov\'{a}%
\ \protect \BOthers {.}}{%
{\protect \APACyear {2018}}%
}]{%
Kolmasova:2018}
\APACinsertmetastar {%
Kolmasova:2018}%
\begin{APACrefauthors}%
Kolma\v{s}ov\'{a}, I.%
, Santol\'{i}k, O.%
, Defer, E.%
, Rison, W.%
, Coquillat, S.%
, Pedeboy, S.%
\BDBL {}Pont, V.%
\end{APACrefauthors}%
\unskip\
\newblock
\APACrefYearMonthDay{2018}{}{}.
\newblock
{\BBOQ}\APACrefatitle {Lightning initiation: Strong pulses of VHF radiation
  accompany preliminary breakdown} {Lightning initiation: Strong pulses of vhf
  radiation accompany preliminary breakdown}.{\BBCQ}
\newblock
\APACjournalVolNumPages{Scientific Reports}{8}{3650}{2045-2322}.
\newblock
\begin{APACrefURL} \url{https://doi.org/10.1038/s41598-018-21972-z}
  \end{APACrefURL}
\newblock
\begin{APACrefDOI} \doi{10.1038/s41598-018-21972-z} \end{APACrefDOI}
\PrintBackRefs{\CurrentBib}

\bibitem [\protect \citeauthoryear {%
Kolma\v{s}ov\'{a}%
\ \protect \BOthers {.}}{%
Kolma\v{s}ov\'{a}%
\ \protect \BOthers {.}}{%
{\protect \APACyear {2014}}%
}]{%
Kolmasova:2014}
\APACinsertmetastar {%
Kolmasova:2014}%
\begin{APACrefauthors}%
Kolma\v{s}ov\'{a}, I.%
, Santolik, O.%
, Farges, T.%
, Rison, W.%
, Lan, R.%
\BCBL {}\ \BBA {} Uhlir, L.%
\end{APACrefauthors}%
\unskip\
\newblock
\APACrefYearMonthDay{2014}{}{}.
\newblock
{\BBOQ}\APACrefatitle {Properties of the unusually short pulse sequences
  occurring prior to the first strokes of negative cloud-to-ground lightning
  flashes} {Properties of the unusually short pulse sequences occurring prior
  to the first strokes of negative cloud-to-ground lightning flashes}.{\BBCQ}
\newblock
\APACjournalVolNumPages{Geophysical Research Letters}{41}{14}{5316-5324}.
\newblock
\begin{APACrefURL}
  \url{https://agupubs.onlinelibrary.wiley.com/doi/abs/10.1002/2014GL060913}
  \end{APACrefURL}
\newblock
\begin{APACrefDOI} \doi{10.1002/2014GL060913} \end{APACrefDOI}
\PrintBackRefs{\CurrentBib}

\bibitem [\protect \citeauthoryear {%
Kostinskiy%
, Marshall%
\BCBL {}\ \BBA {} Stolzenburg%
}{%
Kostinskiy%
\ \protect \BOthers {.}}{%
{\protect \APACyear {2019}}%
}]{%
Kostinskiy:2019}
\APACinsertmetastar {%
Kostinskiy:2019}%
\begin{APACrefauthors}%
Kostinskiy, A\BPBI Y.%
, Marshall, T\BPBI C.%
\BCBL {}\ \BBA {} Stolzenburg, M.%
\end{APACrefauthors}%
\unskip\
\newblock
\APACrefYearMonthDay{2019}{}{}.
\newblock
{\BBOQ}\APACrefatitle {The Mechanism of the Origin and Development of Lightning
  from Initiating Event to Initial Breakdown Pulses} {The mechanism of the
  origin and development of lightning from initiating event to initial
  breakdown pulses}.{\BBCQ}
\newblock
\APACjournalVolNumPages{}{}{arXiv:1906.01033}{}.
\newblock
\begin{APACrefURL} \url{https://arxiv.org/abs/1906.01033} \end{APACrefURL}
\PrintBackRefs{\CurrentBib}

\bibitem [\protect \citeauthoryear {%
{Krehbiel}%
}{%
{Krehbiel}%
}{%
{\protect \APACyear {2017}}%
}]{%
Krehbiel:2017}
\APACinsertmetastar {%
Krehbiel:2017}%
\begin{APACrefauthors}%
{Krehbiel}, P.%
\end{APACrefauthors}%
\unskip\
\newblock
\APACrefYearMonthDay{2017}{Aug}{}.
\newblock
{\BBOQ}\APACrefatitle {Studies of lightning initiation} {Studies of lightning
  initiation}.{\BBCQ}
\newblock
\BIn{} \APACrefbtitle {2017 XXXIInd General Assembly and Scientific Symposium
  of the International Union of Radio Science (URSI GASS)} {2017 xxxiind
  general assembly and scientific symposium of the international union of radio
  science (ursi gass)}\ (\BPG~1-2).
\newblock
\begin{APACrefDOI} \doi{10.23919/URSIGASS.2017.8105171} \end{APACrefDOI}
\PrintBackRefs{\CurrentBib}

\bibitem [\protect \citeauthoryear {%
Lyu%
, Cummer%
, Lu%
, Zhou%
\BCBL {}\ \BBA {} Weinert%
}{%
Lyu%
\ \protect \BOthers {.}}{%
{\protect \APACyear {2016}}%
}]{%
Lyu:2016}
\APACinsertmetastar {%
Lyu:2016}%
\begin{APACrefauthors}%
Lyu, F.%
, Cummer, S\BPBI A.%
, Lu, G.%
, Zhou, X.%
\BCBL {}\ \BBA {} Weinert, J.%
\end{APACrefauthors}%
\unskip\
\newblock
\APACrefYearMonthDay{2016}{}{}.
\newblock
{\BBOQ}\APACrefatitle {Imaging lightning intracloud initial stepped leaders by
  low-frequency interferometric lightning mapping array} {Imaging lightning
  intracloud initial stepped leaders by low-frequency interferometric lightning
  mapping array}.{\BBCQ}
\newblock
\APACjournalVolNumPages{Geophysical Research Letters}{43}{10}{5516-5523}.
\newblock
\begin{APACrefDOI} \doi{10.1002/2016GL069267} \end{APACrefDOI}
\PrintBackRefs{\CurrentBib}

\bibitem [\protect \citeauthoryear {%
Lyu%
, Cummer%
, Qin%
\BCBL {}\ \BBA {} Chen%
}{%
Lyu%
\ \protect \BOthers {.}}{%
{\protect \APACyear {2019}}%
}]{%
Lyu:2019}
\APACinsertmetastar {%
Lyu:2019}%
\begin{APACrefauthors}%
Lyu, F.%
, Cummer, S\BPBI A.%
, Qin, Z.%
\BCBL {}\ \BBA {} Chen, M.%
\end{APACrefauthors}%
\unskip\
\newblock
\APACrefYearMonthDay{2019}{}{}.
\newblock
{\BBOQ}\APACrefatitle {Lightning Initiation Processes Imaged With Very High
  Frequency Broadband Interferometry} {Lightning initiation processes imaged
  with very high frequency broadband interferometry}.{\BBCQ}
\newblock
\APACjournalVolNumPages{Journal of Geophysical Research:
  Atmospheres}{124}{6}{2994-3004}.
\newblock
\begin{APACrefURL}
  \url{https://agupubs.onlinelibrary.wiley.com/doi/abs/10.1029/2018JD029817}
  \end{APACrefURL}
\newblock
\begin{APACrefDOI} \doi{10.1029/2018JD029817} \end{APACrefDOI}
\PrintBackRefs{\CurrentBib}

\bibitem [\protect \citeauthoryear {%
Mareev%
\ \BBA {} Dementyeva%
}{%
Mareev%
\ \BBA {} Dementyeva%
}{%
{\protect \APACyear {2017}}%
}]{%
Mareev:2017}
\APACinsertmetastar {%
Mareev:2017}%
\begin{APACrefauthors}%
Mareev, E\BPBI A.%
\BCBT {}\ \BBA {} Dementyeva, S\BPBI O.%
\end{APACrefauthors}%
\unskip\
\newblock
\APACrefYearMonthDay{2017}{}{}.
\newblock
{\BBOQ}\APACrefatitle {The role of turbulence in thunderstorm, snowstorm, and
  dust storm electrification} {The role of turbulence in thunderstorm,
  snowstorm, and dust storm electrification}.{\BBCQ}
\newblock
\APACjournalVolNumPages{Journal of Geophysical Research:
  Atmospheres}{122}{13}{6976-6988}.
\newblock
\begin{APACrefURL}
  \url{https://agupubs.onlinelibrary.wiley.com/doi/abs/10.1002/2016JD026150}
  \end{APACrefURL}
\newblock
\begin{APACrefDOI} \doi{10.1002/2016JD026150} \end{APACrefDOI}
\PrintBackRefs{\CurrentBib}

\bibitem [\protect \citeauthoryear {%
Marshall%
\ \protect \BOthers {.}}{%
Marshall%
\ \protect \BOthers {.}}{%
{\protect \APACyear {2019}}%
}]{%
Marshall:2019}
\APACinsertmetastar {%
Marshall:2019}%
\begin{APACrefauthors}%
Marshall, T.%
, Bandara, S.%
, Karunarathne, N.%
, Karunarathne, S.%
, Kolmasova, I.%
, Siedlecki, R.%
\BCBL {}\ \BBA {} Stolzenburg, M.%
\end{APACrefauthors}%
\unskip\
\newblock
\APACrefYearMonthDay{2019}{}{}.
\newblock
{\BBOQ}\APACrefatitle {A study of lightning flash initiation prior to the first
  initial breakdown pulse} {A study of lightning flash initiation prior to the
  first initial breakdown pulse}.{\BBCQ}
\newblock
\APACjournalVolNumPages{Atmospheric Research}{217}{}{10 - 23}.
\newblock
\begin{APACrefURL}
  \url{http://www.sciencedirect.com/science/article/pii/S0169809518307555}
  \end{APACrefURL}
\newblock
\begin{APACrefDOI} \doi{https://doi.org/10.1016/j.atmosres.2018.10.013}
  \end{APACrefDOI}
\PrintBackRefs{\CurrentBib}

\bibitem [\protect \citeauthoryear {%
Montany\`{a}%
, van~der Velde%
\BCBL {}\ \BBA {} Williams%
}{%
Montany\`{a}%
\ \protect \BOthers {.}}{%
{\protect \APACyear {2015}}%
}]{%
Montanya:2015}
\APACinsertmetastar {%
Montanya:2015}%
\begin{APACrefauthors}%
Montany\`{a}, J.%
, van~der Velde, O.%
\BCBL {}\ \BBA {} Williams, E\BPBI R.%
\end{APACrefauthors}%
\unskip\
\newblock
\APACrefYearMonthDay{2015}{}{}.
\newblock
{\BBOQ}\APACrefatitle {The start of lightning: Evidence of bidirectional
  lightning initiation} {The start of lightning: Evidence of bidirectional
  lightning initiation}.{\BBCQ}
\newblock
\APACjournalVolNumPages{Scientific Reports}{5}{1}{15180}.
\newblock
\begin{APACrefURL} \url{https://doi.org/10.1038/srep15180} \end{APACrefURL}
\newblock
\begin{APACrefDOI} \doi{10.1038/srep15180} \end{APACrefDOI}
\PrintBackRefs{\CurrentBib}

\bibitem [\protect \citeauthoryear {%
Mulrey%
\ \protect \BOthers {.}}{%
Mulrey%
\ \protect \BOthers {.}}{%
{\protect \APACyear {2019}}%
}]{%
Mulrey:2019}
\APACinsertmetastar {%
Mulrey:2019}%
\begin{APACrefauthors}%
Mulrey, K.%
, Bonardi, A.%
, Buitink, S.%
, Corstanje, A.%
, Falcke, H.%
, Hare, B.%
\BDBL {}Winchen, T.%
\end{APACrefauthors}%
\unskip\
\newblock
\APACrefYearMonthDay{2019}{}{}.
\newblock
{\BBOQ}\APACrefatitle {Calibration of the LOFAR low-band antennas using the
  Galaxy and a model of the signal chain} {Calibration of the lofar low-band
  antennas using the galaxy and a model of the signal chain}.{\BBCQ}
\newblock
\APACjournalVolNumPages{Astroparticle Physics}{111}{}{1 - 11}.
\newblock
\begin{APACrefURL}
  \url{http://www.sciencedirect.com/science/article/pii/S0927650518302810}
  \end{APACrefURL}
\newblock
\begin{APACrefDOI} \doi{https://doi.org/10.1016/j.astropartphys.2019.03.004}
  \end{APACrefDOI}
\PrintBackRefs{\CurrentBib}

\bibitem [\protect \citeauthoryear {%
Pel%
}{%
Pel%
}{%
{\protect \APACyear {2019}}%
}]{%
Pel:2019}
\APACinsertmetastar {%
Pel:2019}%
\begin{APACrefauthors}%
Pel, A.%
\end{APACrefauthors}%
\unskip\
\newblock
\APACrefYear{2019}.
\unskip\
\newblock
\APACrefbtitle {Imaging Lightning with the Extended Kalman Filter} {Imaging
  lightning with the extended kalman filter}\ \APACtypeAddressSchool {Master
  Thesis}{KVI-CART}{University of Groningen, Faculty of Science and
  Engeneering}.
\unskip\
\newblock
\begin{APACrefURL} \url{http://fse.studenttheses.ub.rug.nl/id/eprint/19751}
  \end{APACrefURL}
\PrintBackRefs{\CurrentBib}

\bibitem [\protect \citeauthoryear {%
Pu%
\ \BBA {} Cummer%
}{%
Pu%
\ \BBA {} Cummer%
}{%
{\protect \APACyear {2019}}%
}]{%
Pu:2019}
\APACinsertmetastar {%
Pu:2019}%
\begin{APACrefauthors}%
Pu, Y.%
\BCBT {}\ \BBA {} Cummer, S\BPBI A.%
\end{APACrefauthors}%
\unskip\
\newblock
\APACrefYearMonthDay{2019}{}{}.
\newblock
{\BBOQ}\APACrefatitle {Needles and Lightning Leader Dynamics Imaged with 100 --
  200 MHz Broadband VHF Interferometry} {Needles and lightning leader dynamics
  imaged with 100 -- 200 mhz broadband vhf interferometry}.{\BBCQ}
\newblock
\APACjournalVolNumPages{Geophysical Research Letters}{46}{22}{13556-13563}.
\newblock
\begin{APACrefURL}
  \url{https://agupubs.onlinelibrary.wiley.com/doi/abs/10.1029/2019GL085635}
  \end{APACrefURL}
\newblock
\begin{APACrefDOI} \doi{10.1029/2019GL085635} \end{APACrefDOI}
\PrintBackRefs{\CurrentBib}

\bibitem [\protect \citeauthoryear {%
Qi%
\ \protect \BOthers {.}}{%
Qi%
\ \protect \BOthers {.}}{%
{\protect \APACyear {2016}}%
}]{%
Qi:2016}
\APACinsertmetastar {%
Qi:2016}%
\begin{APACrefauthors}%
Qi, Q.%
, Lu, W.%
, Ma, Y.%
, Chen, L.%
, Zhang, Y.%
\BCBL {}\ \BBA {} Rakov, V\BPBI A.%
\end{APACrefauthors}%
\unskip\
\newblock
\APACrefYearMonthDay{2016}{}{}.
\newblock
{\BBOQ}\APACrefatitle {High-speed video observations of the fine structure of a
  natural negative stepped leader at close distance} {High-speed video
  observations of the fine structure of a natural negative stepped leader at
  close distance}.{\BBCQ}
\newblock
\APACjournalVolNumPages{Atmospheric Research}{178-179}{}{260 - 267}.
\newblock
\begin{APACrefDOI} \doi{https://doi.org/10.1016/j.atmosres.2016.03.027}
  \end{APACrefDOI}
\PrintBackRefs{\CurrentBib}

\bibitem [\protect \citeauthoryear {%
Rhodes%
, Shao%
, Krehbiel%
, Thomas%
\BCBL {}\ \BBA {} Hayenga%
}{%
Rhodes%
\ \protect \BOthers {.}}{%
{\protect \APACyear {1994}}%
}]{%
Rhodes:1994}
\APACinsertmetastar {%
Rhodes:1994}%
\begin{APACrefauthors}%
Rhodes, C\BPBI T.%
, Shao, X\BPBI M.%
, Krehbiel, P\BPBI R.%
, Thomas, R\BPBI J.%
\BCBL {}\ \BBA {} Hayenga, C\BPBI O.%
\end{APACrefauthors}%
\unskip\
\newblock
\APACrefYearMonthDay{1994}{}{}.
\newblock
{\BBOQ}\APACrefatitle {Observations of lightning phenomena using radio
  interferometry} {Observations of lightning phenomena using radio
  interferometry}.{\BBCQ}
\newblock
\APACjournalVolNumPages{Journal of Geophysical Research:
  Atmospheres}{99}{D6}{13059-13082}.
\newblock
\begin{APACrefDOI} \doi{10.1029/94JD00318} \end{APACrefDOI}
\PrintBackRefs{\CurrentBib}

\bibitem [\protect \citeauthoryear {%
Rison%
\ \protect \BOthers {.}}{%
Rison%
\ \protect \BOthers {.}}{%
{\protect \APACyear {2016}}%
}]{%
Rison:2016}
\APACinsertmetastar {%
Rison:2016}%
\begin{APACrefauthors}%
Rison, W.%
, Krehbiel, P\BPBI R.%
, Stock, M\BPBI G.%
, Edens, H\BPBI E.%
, Shao, X\BHBI M.%
, Thomas, R\BPBI J.%
\BDBL {}Zhang, Y.%
\end{APACrefauthors}%
\unskip\
\newblock
\APACrefYearMonthDay{2016}{}{}.
\newblock
{\BBOQ}\APACrefatitle {Observations of narrow bipolar events reveal how
  lightning is initiated in thunderstorms} {Observations of narrow bipolar
  events reveal how lightning is initiated in thunderstorms}.{\BBCQ}
\newblock
\APACjournalVolNumPages{Nature Communications}{7}{1}{10721}.
\newblock
\begin{APACrefURL} \url{https://doi.org/10.1038/ncomms10721} \end{APACrefURL}
\newblock
\begin{APACrefDOI} \doi{10.1038/ncomms10721} \end{APACrefDOI}
\PrintBackRefs{\CurrentBib}

\bibitem [\protect \citeauthoryear {%
Rison%
, Thomas%
, Krehbiel%
, Hamlin%
\BCBL {}\ \BBA {} Harlin%
}{%
Rison%
\ \protect \BOthers {.}}{%
{\protect \APACyear {1999}}%
}]{%
Rison:1999}
\APACinsertmetastar {%
Rison:1999}%
\begin{APACrefauthors}%
Rison, W.%
, Thomas, R\BPBI J.%
, Krehbiel, P\BPBI R.%
, Hamlin, T.%
\BCBL {}\ \BBA {} Harlin, J.%
\end{APACrefauthors}%
\unskip\
\newblock
\APACrefYearMonthDay{1999}{}{}.
\newblock
{\BBOQ}\APACrefatitle {A GPS-based three-dimensional lightning mapping system:
  Initial observations in central New Mexico} {A gps-based three-dimensional
  lightning mapping system: Initial observations in central new mexico}.{\BBCQ}
\newblock
\APACjournalVolNumPages{Geophysical Research Letters}{26}{23}{3573-3576}.
\newblock
\begin{APACrefDOI} \doi{10.1029/1999GL010856} \end{APACrefDOI}
\PrintBackRefs{\CurrentBib}

\bibitem [\protect \citeauthoryear {%
Rutjes%
, Ebert%
, Buitink%
, Scholten%
\BCBL {}\ \BBA {} Trinh%
}{%
Rutjes%
\ \protect \BOthers {.}}{%
{\protect \APACyear {2019}}%
}]{%
Rutjes:2019}
\APACinsertmetastar {%
Rutjes:2019}%
\begin{APACrefauthors}%
Rutjes, C.%
, Ebert, U.%
, Buitink, S.%
, Scholten, O.%
\BCBL {}\ \BBA {} Trinh, T\BPBI N.%
\end{APACrefauthors}%
\unskip\
\newblock
\APACrefYearMonthDay{2019}{}{}.
\newblock
{\BBOQ}\APACrefatitle {Generation of Seed Electrons by Extensive Air Showers,
  and the Lightning Inception Problem Including Narrow Bipolar Events}
  {Generation of seed electrons by extensive air showers, and the lightning
  inception problem including narrow bipolar events}.{\BBCQ}
\newblock
\APACjournalVolNumPages{Journal of Geophysical Research:
  Atmospheres}{124}{13}{7255-7269}.
\newblock
\begin{APACrefURL}
  \url{https://agupubs.onlinelibrary.wiley.com/doi/abs/10.1029/2018JD029040}
  \end{APACrefURL}
\newblock
\begin{APACrefDOI} \doi{10.1029/2018JD029040} \end{APACrefDOI}
\PrintBackRefs{\CurrentBib}

\bibitem [\protect \citeauthoryear {%
Scholten%
}{%
Scholten%
}{%
{\protect \APACyear {2020}}%
}]{%
Scholten_LofarImaging:2020}
\APACinsertmetastar {%
Scholten_LofarImaging:2020}%
\begin{APACrefauthors}%
Scholten, O.%
\end{APACrefauthors}%
\unskip\
\newblock
\APACrefYearMonthDay{2020}{}{}.
\newblock
\APACrefbtitle {A practical guide to Lightning Imaging with LOFAR} {A practical
  guide to lightning imaging with lofar}\ \APACbVolEdTR {}{internal report}.
\newblock
\APACaddressInstitution{}{Kapteyn Institute, University of Groningen, NL}.
\newblock
\begin{APACrefURL} \url{https://www.astro.rug.nl/~scholten/Lightning/LOFLI.htm}
  \end{APACrefURL}
\PrintBackRefs{\CurrentBib}

\bibitem [\protect \citeauthoryear {%
Scholten%
\ \protect \BOthers {.}}{%
Scholten%
\ \protect \BOthers {.}}{%
{\protect \APACyear {2020}}%
}]{%
Scholten:BBLofar}
\APACinsertmetastar {%
Scholten:BBLofar}%
\begin{APACrefauthors}%
Scholten, O.%
\BCBT {}\ \BOthersPeriod {.}
\end{APACrefauthors}%
\unskip\
\newblock
\APACrefYearMonthDay{2020}{}{}.
\newblock
{\BBOQ}\APACrefatitle {Broadband radio pulses and LOFAR imaging} {Broadband
  radio pulses and lofar imaging}.{\BBCQ}
\newblock
\APACjournalVolNumPages{to be submitted to JGR}{}{}{}.
\PrintBackRefs{\CurrentBib}

\bibitem [\protect \citeauthoryear {%
Stock%
\ \protect \BOthers {.}}{%
Stock%
\ \protect \BOthers {.}}{%
{\protect \APACyear {2014}}%
}]{%
Stock:2014}
\APACinsertmetastar {%
Stock:2014}%
\begin{APACrefauthors}%
Stock, M\BPBI G.%
, Akita, M.%
, Krehbiel, P\BPBI R.%
, Rison, W.%
, Edens, H\BPBI E.%
, Kawasaki, Z.%
\BCBL {}\ \BBA {} Stanley, M\BPBI A.%
\end{APACrefauthors}%
\unskip\
\newblock
\APACrefYearMonthDay{2014}{}{}.
\newblock
{\BBOQ}\APACrefatitle {Continuous broadband digital interferometry of lightning
  using a generalized cross-correlation algorithm} {Continuous broadband
  digital interferometry of lightning using a generalized cross-correlation
  algorithm}.{\BBCQ}
\newblock
\APACjournalVolNumPages{Journal of Geophysical Research:
  Atmospheres}{119}{6}{3134-3165}.
\newblock
\begin{APACrefDOI} \doi{10.1002/2013JD020217} \end{APACrefDOI}
\PrintBackRefs{\CurrentBib}

\bibitem [\protect \citeauthoryear {%
Stolzenburg%
, Marshall%
\BCBL {}\ \BBA {} Karunarathne%
}{%
Stolzenburg%
\ \protect \BOthers {.}}{%
{\protect \APACyear {2020}}%
}]{%
Stolzenburg:2020}
\APACinsertmetastar {%
Stolzenburg:2020}%
\begin{APACrefauthors}%
Stolzenburg, M.%
, Marshall, T\BPBI C.%
\BCBL {}\ \BBA {} Karunarathne, S.%
\end{APACrefauthors}%
\unskip\
\newblock
\APACrefYearMonthDay{2020}{}{}.
\newblock
{\BBOQ}\APACrefatitle {On the Transition From Initial Leader to Stepped Leader
  in Negative Cloud-to-Ground Lightning} {On the transition from initial leader
  to stepped leader in negative cloud-to-ground lightning}.{\BBCQ}
\newblock
\APACjournalVolNumPages{Journal of Geophysical Research:
  Atmospheres}{125}{4}{e2019JD031765}.
\newblock
\begin{APACrefURL}
  \url{https://agupubs.onlinelibrary.wiley.com/doi/abs/10.1029/2019JD031765}
  \end{APACrefURL}
\newblock
\APACrefnote{e2019JD031765 2019JD031765}
\newblock
\begin{APACrefDOI} \doi{10.1029/2019JD031765} \end{APACrefDOI}
\PrintBackRefs{\CurrentBib}

\bibitem [\protect \citeauthoryear {%
Tran%
\ \BBA {} Rakov%
}{%
Tran%
\ \BBA {} Rakov%
}{%
{\protect \APACyear {2016}}%
}]{%
Tran:2016}
\APACinsertmetastar {%
Tran:2016}%
\begin{APACrefauthors}%
Tran, M\BPBI D.%
\BCBT {}\ \BBA {} Rakov, V\BPBI A.%
\end{APACrefauthors}%
\unskip\
\newblock
\APACrefYearMonthDay{2016}{}{}.
\newblock
{\BBOQ}\APACrefatitle {Initiation and propagation of cloud-to-ground lightning
  observed with a high-speed video camera} {Initiation and propagation of
  cloud-to-ground lightning observed with a high-speed video camera}.{\BBCQ}
\newblock
\APACjournalVolNumPages{Scientific Reports}{6}{1}{39521}.
\newblock
\begin{APACrefURL} \url{https://doi.org/10.1038/srep39521} \end{APACrefURL}
\newblock
\begin{APACrefDOI} \doi{10.1038/srep39521} \end{APACrefDOI}
\PrintBackRefs{\CurrentBib}

\bibitem [\protect \citeauthoryear {%
Trinh%
\ \protect \BOthers {.}}{%
Trinh%
\ \protect \BOthers {.}}{%
{\protect \APACyear {2020}}%
}]{%
Trinh:2020}
\APACinsertmetastar {%
Trinh:2020}%
\begin{APACrefauthors}%
Trinh, T\BPBI N\BPBI G.%
, Scholten, O.%
, Buitink, S.%
, Ebert, U.%
, Hare, B\BPBI M.%
, Krehbiel, P\BPBI R.%
\BDBL {}Winchen, T.%
\end{APACrefauthors}%
\unskip\
\newblock
\APACrefYearMonthDay{2020}{}{}.
\newblock
{\BBOQ}\APACrefatitle {Determining Electric Fields in Thunderclouds With the
  Radiotelescope LOFAR} {Determining electric fields in thunderclouds with the
  radiotelescope lofar}.{\BBCQ}
\newblock
\APACjournalVolNumPages{Journal of Geophysical Research:
  Atmospheres}{125}{8}{e2019JD031433}.
\newblock
\begin{APACrefURL}
  \url{https://agupubs.onlinelibrary.wiley.com/doi/abs/10.1029/2019JD031433}
  \end{APACrefURL}
\newblock
\APACrefnote{e2019JD031433 10.1029/2019JD031433}
\newblock
\begin{APACrefDOI} \doi{10.1029/2019JD031433} \end{APACrefDOI}
\PrintBackRefs{\CurrentBib}

\bibitem [\protect \citeauthoryear {%
van Haarlem%
\ \protect \BOthers {.}}{%
van Haarlem%
\ \protect \BOthers {.}}{%
{\protect \APACyear {2013}}%
}]{%
Haarlem:2013}
\APACinsertmetastar {%
Haarlem:2013}%
\begin{APACrefauthors}%
van Haarlem, M\BPBI P.%
\BCBT {}\ \BOthersPeriod {.}
\end{APACrefauthors}%
\unskip\
\newblock
\APACrefYearMonthDay{2013}{}{}.
\newblock
{\BBOQ}\APACrefatitle {{LOFAR: The LOw-Frequency ARray}} {{LOFAR: The
  LOw-Frequency ARray}}.{\BBCQ}
\newblock
\APACjournalVolNumPages{A\&A}{556}{}{A2}.
\newblock
\begin{APACrefDOI} \doi{10.1051/0004-6361/201220873} \end{APACrefDOI}
\PrintBackRefs{\CurrentBib}

\bibitem [\protect \citeauthoryear {%
Yoshida%
\ \protect \BOthers {.}}{%
Yoshida%
\ \protect \BOthers {.}}{%
{\protect \APACyear {2010}}%
}]{%
Yoshida:2010}
\APACinsertmetastar {%
Yoshida:2010}%
\begin{APACrefauthors}%
Yoshida, S.%
, Biagi, C\BPBI J.%
, Rakov, V\BPBI A.%
, Hill, J\BPBI D.%
, Stapleton, M\BPBI V.%
, Jordan, D\BPBI M.%
\BDBL {}Kawasaki, Z\BHBI I.%
\end{APACrefauthors}%
\unskip\
\newblock
\APACrefYearMonthDay{2010}{}{}.
\newblock
{\BBOQ}\APACrefatitle {Three-dimensional imaging of upward positive leaders in
  triggered lightning using VHF broadband digital interferometers}
  {Three-dimensional imaging of upward positive leaders in triggered lightning
  using vhf broadband digital interferometers}.{\BBCQ}
\newblock
\APACjournalVolNumPages{Geophysical Research Letters}{37}{5}{}.
\newblock
\begin{APACrefDOI} \doi{10.1029/2009GL042065} \end{APACrefDOI}
\PrintBackRefs{\CurrentBib}

\end{thebibliography}

\end{document}